\documentclass[10pt,letterpaper]{article}
\usepackage[top=0.85in,left=1.55in,footskip=0.75in]{geometry}

\usepackage{amsmath,amssymb}

\usepackage{changepage}

\usepackage{textcomp,marvosym}

\usepackage{cite}

\usepackage{nameref,hyperref}

\usepackage[right]{lineno}

\usepackage[nopatch=eqnum]{microtype}
\DisableLigatures[f]{encoding = *, family = * }

\usepackage[table]{xcolor}

\usepackage{array}

\newcolumntype{+}{!{\vrule width 2pt}}

\newlength\savedwidth

\usepackage[aboveskip=1pt,labelfont=bf,labelsep=period,justification=raggedright,singlelinecheck=off]{caption}

\makeatletter
\renewcommand{\@biblabel}[1]{\quad#1.}
\makeatother

\usepackage{longtable}
\usepackage{float}
\usepackage{lastpage,fancyhdr,graphicx}
\usepackage{epstopdf}
\fancyheadoffset[L]{2.25in}
\fancyfootoffset[L]{2.25in}
\begin{document}
\vspace*{0.2in}

\begin{flushleft}
{\Large
\textbf\newline{Modelling human activities in a system of cities} % Please use "sentence case" for title and headings (capitalize only the first word in a title (or heading), the first word in a subtitle (or subheading), and any proper nouns).
}
\newline
\\
Guo-Shiuan Lin\textsuperscript{1},
Denise Hertwig\textsuperscript{2},
Megan McGrory\textsuperscript{2},
Tiancheng Ma\textsuperscript{2},
Stefán Thor Smith\textsuperscript{3},
Maider Llaguno-Munitxa\textsuperscript{4}, 
Sue Grimmond\textsuperscript{2},
Gabriele Manoli\textsuperscript{1,*}
\\
\bigskip
\textbf{1} Laboratory of Urban and Environmental Systems, School of Architecture, Civil and Environmental Engineering, École Polytechnique Fédérale de Lausanne, Switzerland
\\
\textbf{2} Department of Meteorology, University of Reading, United Kingdom
\\
\textbf{3} School of the Built Environment, University of Reading, United Kingdom
\\
\textbf{4} Louvain Research Institute of Landscape, Architecture, Built Environment, Université Catholique de Louvain (UCLouvain), Ottignies-Louvain-la-Neuve 1348, Belgium
% \bigskip

% Insert additional author notes using the symbols described below. Insert symbol callouts after author names as necessary.
% 
% Remove or comment out the author notes below if they aren't used.
%
% Primary Equal Contribution Note
% \Yinyang These authors contributed equally to this work.

% Additional Equal Contribution Note
% Also use this double-dagger symbol for special authorship notes, such as senior authorship.
% \ddag These authors also contributed equally to this work.

% \textcurrency Current Address: Dept/Program/Center, Institution Name, City, State, Country % change symbol to "\textcurrency a" if more than one current address note
% \textcurrency b Insert second current address 
% \textcurrency c Insert third current address

% Use the asterisk to denote corresponding authorship and provide email address in note below.
* gabriele.manoli@epfl.ch

\end{flushleft}
\section*{Abstract}
Cities host most of the world population with diverse services and activities. One key challenge in urban modelling is the quantification of intra- and inter-city mobility patterns and the associated space-time dynamics of population density and anthropogenic activities. To address this, we apply the novel agent-based urban model DAVE (Dynamic Anthropogenic actiVities and feedback to Emissions) to simulate population behaviour and mobility in the Vaud and Geneva Cantons, a system of small- to medium-size cities in Switzerland. Simulation results provide detailed temporal (10 min) and spatial (500 m) population dynamics for different age groups and day types. DAVE further models the time-varying population distribution in 11 different microenvironments (e.g., home, work, leisure, outdoor) and the travel flows by different modes. Simulation results align with observations, confirming the possibility of driving urban system modelling with statistical information on residents' behaviour. Sustainability and health indicators like daily driving distance and walking time for each neighbourhood are also reflected by the model with urban-rural gradients displayed. This work serves as a foundation for future applications of DAVE to study bottom-up human-built environment interactions, from anthropogenic emissions and building energy to urban climate, exposure, and health in cities around the world.

% \linenumbers
\section*{Introduction}
Cities are characterised by high population densities and intense human activities as more than half of the global population now lives in urban areas \cite{grimm_global_2008}. While being rich in economic opportunities, attractive amenities, and transport connectivity \cite{bettencourt_growth_2007, strano_elementary_2012}, cities also suffer from negative anthropogenic impacts such as traffic congestion \cite{lu_expansion_2021, xu_road_2024}, air pollution, intensification of extreme heat \cite{gao_urbanization-induced_2024, jiang_amplified_2019}, and disease transmission \cite{mazza_data-driven_2025, brizuela_understanding_2021}. To reduce these negative phenomena and design healthier and more sustainable urban systems, it is crucial to understand how humans and their daily activities interact with the form and functions of cities \cite{barlow_developing_2017}. As studies have shown, human behaviour can modify urban environments and their climate over both short and long timescales. These influences range from decadal urban development (e.g., \cite{hendrick_dynamic_2025,capel-timms_angiogenic_2024 ,li_divergent_2023, ullah_impact_2023}), to seasonal energy consumption (e.g., \cite{offerle_temporal_2006, ichinose_impact_1999}), daily traffic fluctuations (e.g., \cite{wu_effects_2021, teufel_impact_2021, bao_does_2020}) and hourly activity patterns (e.g., \cite{offerle_temporal_2006, sailor_topdown_2004}), which affect anthropogenic emissions of heat, carbon, and pollutants (e.g., \cite{lu_expansion_2021, vahmani_anthropogenic_2022, glaeser_greenness_2010}).

In addition to observations, models have been developed to account for such human--urban environment interactions and enable climate simulations at different scales, locations, and scenarios \cite{lipson_evaluation_2024, skamarock_description_nodate}. However, many models only account for certain sources of anthropogenic emissions such as air conditioning \cite{salamanca_new_2010}, or follow top-down methods that rely on existent inventories to approximate emissions (e.g., \cite{sailor_topdown_2004,lu_estimate_2016}). Although efficient to implement and update, such models may overlook spatial and temporal variations \cite{swan_modeling_2009}, and neglect the complex feedback mechanisms governing urban multi-sector dynamics. With the advancement in computation power and data availability, bottom-up models, which relate individual end-uses towards their aggregated effects on urban systems are increasingly being used. Agent-based models hence emerge as powerful tools to simulate the heterogeneous behaviours of diverse populations within complex urban systems (e.g., \cite{bastarianto_agent-based_2023, vuthi_agent-based_2022, chen_agent-based_2012}). Agent-based models allow independent autonomous entities (agents) to interact with each other and their environment by following deterministic or probabilistic rules, leading to replication of some observed system-level behaviours. As such, agent-based models have been developed for anthropogenic heat emissions \cite{capel-timms_dynamic_2020, schoetter_parametrisation_2017}, building energy use \cite{wang_cesar_2018, micolier_li-bim_2019}, and human mobility \cite{w_axhausen_multi-agent_2016, auld_polaris_2016, saprykin_accelerating_2022}. However, these models are often sector-specific and decoupled from the urban climate and other sectors in the urban system. To summarize, despite advances in urban modelling in recent years, existing approaches suffer from two main limitations: they are often static and/or focus on single urban sectors \cite{wilson_future_2018}. There is still a lack of system-level approaches that integrate all relevant facets of a city to fully capture the interactions and feedback processes between them.

%Introduction of DAVE
A notable exception is the newly developed model DAVE (Dynamic Anthropogenic actiVities and feedback to Emissions). DAVE is a multi-scale agent-based model that is capable of simulating human behaviour across spatial scales from buildings to neighbourhoods and cities \cite{hertwig_dave_2025, hertwig_connecting_2025}. Evolving from DASH \cite{capel-timms_dynamic_2020}, DAVE uses observed time-use behaviour of different demographic groups, together with geospatial data, to simulate (1) activity schedules of different socioeconomic groups, (2) dynamic population distribution, (3) building occupancy and anthropogenic heat emissions, and (4) indoor and outdoor climate variables. DAVE has been applied to the cities of London \cite{hertwig_dave_2025, hertwig_connecting_2025} and Berlin \cite{hertwig_modelling_2024} but its ability and transferability to simulate the aforementioned dynamics in a system of cities \cite{berry_cities_1964, pred_city-systems_2017} remains to be tested. 

Cities are rarely isolated entities. Rather, they function within a broader network of urban settlements where mobility, trade, cooperation, and competition shape their development \cite{crosato_polycentric_2021,ellam_stochastic_2018, monechi_hamiltonian_2020} and the high-frequency dynamics of their daily activities \cite{batty_visualizing_2018}. At the regional scale, the major linkage between different city centres and the surrounding rural areas is represented by urban mobility systems (from roads to public transport, railways and active mobility paths), with hundreds of millions of people routinely commuting every day in the US and Europe\cite{burrows_commuting_2024, eurostat_main_2020}. As such, inter-city activities and mobility are tightly connected, which is, however, often overlooked by current models. Improving modelling on the inter-city spatial interaction can unravel how activity decisions and social processes change from a single city to a multi-city system. This allows for addressing issues across city boundaries and facilitates regional planning and coordination. Only a modelling framework that spans multiple scales of interest (from individuals to systems of cities) can provide a holistic understanding of urban systems and clarify how microscopic-level dynamics give rise to emergent macroscopic patterns -- aspects that are not yet fully understood \cite{hendrick_stochastic_2025}.

% what we did
Here we address this challenge based on the example of the neighbouring Swiss cantons Vaud and Geneva. This region consists of two major metropolitan areas in Switzerland (Lausanne and Geneva), surrounded by rural areas and small-size cities and villages connected by extensive road and public transport networks. Thus, it provides an ideal setting for testing the transferability of DAVE to another country and at the inter-city scale for the first time. We focus on DAVE's behaviour and transport modules and obtain individual-level activity-based mesoscopic mobility simulations for about 1 million citizens at 500-m and 10-min resolution. The results provide critical information on spatially and temporally varying population densities and activities, which will serve as a basis for future multi-city bottom-up modelling, from building energy applications, to urban climate, and environmental risk assessments.

\section*{Materials and methods}
\subsection*{Study domain and period}
The study domain covers Canton Vaud and Canton Geneva in Western Switzerland with a population of around 1.3 million people (of which about 1 million are older than 18 and are included in the simulations, referred to as `citizens'). The area is composed of small- to medium-size cities, among which Geneva and Lausanne are the biggest followed by Yverdon-les-bains and Montreux, and a large rural and alpine area characterized by a complex topography and the presence of the Leman lake. The main cities are well-connected with roads (both highways and local roads) and public transport networks (particularly trains). The car ownership is between 42 \% (Geneva) and 52 \% (Vaud), slightly lower than the national average (54 \%)(\cite{noauthor_level_2025}), reflecting the stronger use of public transport and a more compact urban setting, especially for Geneva. There are strong work-related commuting flows by both private and public transport between Lausanne and Geneva as well as from rural regions towards the major city centres. As such, this area offers an ideal setting to investigate human mobility and activities in a system of inter-connected urban settlements. For the DAVE simulation, the domain is divided into 13,243 model grid cells with a spatial resolution of 500~m, called `neighbourhoods.' The simulation was run at 10-min resolution for 16-19 July 2022 (Saturday to Tuesday) when there was a major heatwave in Switzerland to allow for future studies on heat exposure. 

\subsection*{Urban system model DAVE}
\begin{figure}[!h]
\includegraphics[width=0.9\textwidth]{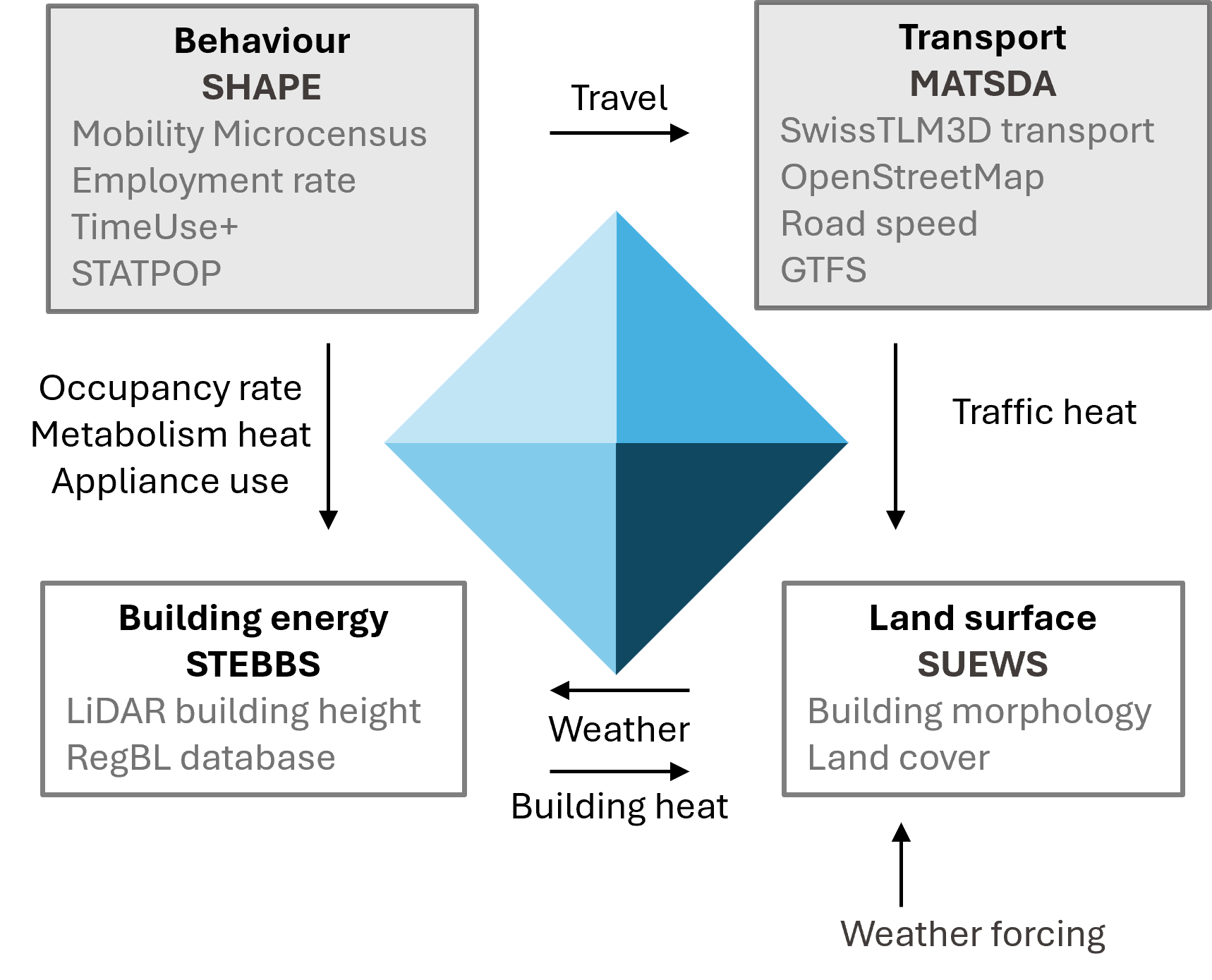} 
\caption{{\bf DAVE modules and the main data used in this study.}
The Behaviour and Transport modules are the focus of this study. The arrows show the main exchanges between modules. Adapted from \cite{hertwig_connecting_2025}.}
\label{fig1}
\end{figure}

DAVE is a multi-scale agent-based model that employs a bottom-up approach to simulate the interactions and feedback between humans and physical urban spaces \cite{hertwig_dave_2025, hertwig_connecting_2025}. The agents in DAVE are the 500~m neighbourhoods which include attributes of citizens, building archetypes, land cover, and transport systems. Agents interact with each other through movements of citizens (population flows) and with the environment through heat exchanges/releases from buildings, transport, and human metabolism. DAVE consists of four inter-dependent modules: behaviour (i.e., the timing and sequence of activities) (SHAPE, Scheduler for Human Activities and energy exPEnditure), transport (MATSDA, Movement And Travel Simulations using Dijkstra’s Algorithm) \cite{ma_transport_2026}, building energy (STEBBS, Simplified Thermal Energy Balance for Building Scheme)\cite{liu_vertically_2026, capel-timms_dynamic_2020}, and land surface (SUEWS, Surface Urban Energy and Water balance Scheme) \cite{grimmond_suews_2025, sun_suews_2019, jarvi_surface_2011}. The main structure of DAVE and the data used for this study are shown in Fig.~\ref{fig1}. The coupling of these modules allows to understand how changes in one component affect other aspects of the urban system. For example, human activities are directly related to the energy use in different indoor microenvironments (MEs), and the energy demand influence the need for building conditioning. Together, the energy consumption in these spaces inform anthropogenic heat emissions, thus altering urban energy fluxes and a neighbourhoods' micro-climate. The citizen objects, in turn, can also dynamically respond to changes in the environment (e.g., traffic disruptions, weather, and socio-economic changes \cite{hertwig_dave_2025}). In this study, we focus on the SHAPE and MATSDA modules to simulate  mobility and activity patterns in the Lausanne-Geneva area (see Fig.~\ref{fig2}). This allows testing the model's transferability to different geographic contexts but also lays the foundation for future applications of the complete DAVE framework to include more Swiss cities. 

\subsection*{Input data preparation}
Multiple input data have to be prepared to run DAVE (details in \cite{hertwig_dave_2025, hertwig_connecting_2025}). For running SHAPE and MATSDA, the following data are the main input requirements: activity profiles, population distribution, microenvironments and spatial attractors, and transport databases. The data generation follows the MAPSECC approach (Multi-scale harmonisation Across Physical and Socio-Economic Characteristics of a City region; \cite{hertwig_connecting_2025, hertwig_multi-scale_2024}). 

\subsubsection*{Activity profiles}
Diurnal activity schedules of different population groups distinguish between weekend or weekdays as well as work (school) and non-work (non-school) days. Activities are linked to different microenvironments (MEs), which are generalised settings where a person spends time doing certain activities \cite{hertwig_connecting_2025}. The activity schedules are represented as a temporal sequence of `activity profiles', segmented by changes in MEs or days. Activity profiles can be derived from time-use surveys (TUS) which record how people use their daily time; for example, the UK TUS is used to run DAVE for London \cite{capel-timms_dynamic_2020, hertwig_connecting_2025}. For the Swiss population, TimeUse+ \cite{winkler_timeuse_2024} is the most recent TUS with the most detailed information on participants' characteristics and activities. TimeUse+ was conducted from July 2022 to February 2023 with activities being recorded by 1,318 participants over four continuous weeks.  While the survey was restricted to German-speaking Switzerland, mobility patterns recorded cover the entire country. Participants could select the survey start date and most started tracking in October and November 2022. They documented their locations (continuously) and activities (10-min resolution) during the survey period through the TimeUse+ App on their mobile phones. The age of participants ranges from 18 to  75 years, covering different economic activity groups (74\% employed, 9\% retired, 10\% students; remainder unspecified) ). The majority of participants belong to the young to middle age group (18--40; 40\%) while only 9\% are older than 65 years. Participants are anonymous and their location is anonymized to 100-m resolution.

TimeUse+ distinguishes two types of events. First, it considers `track events' when participants are on the move. Their locations and transport modes were passively tracked/detected via the TimeUse+ App with the possibility of manual correction of the transport modes by the participants. Second, `stay events' identify periods when participants remained at a fixed location. Stay events are marked by participants as `Home', `Work', or `Other' to indicate the environment in which they are located. For each stay event, a single or multiple activities can occur. Participants had to manually choose the activities they performed at the location from a predefined activity list (including 28 activities such as working, sleeping, shopping, exercising, cooking, etc., full list given by Winkler et al. \cite{winkler_timeuse_2024}). They are also asked to record the duration of each activity and ensure the sum equals the total duration of the stay event. With non-stop tracking for four weeks, the diurnal and weekly behaviour profiles provide representation of timing of activity types in the Swiss population.

The locations and activities in TimeUse+ are mapped to 11 MEs (mostly following the definition in \cite{hertwig_connecting_2025}): Home, Work, Shop, Hospitality, Indoor Leisure, Outdoor, Healthcare, University, Hotel, Other Home, Primary School, and Secondary School. Together they constitute the essential urban social functions to support urban livability defined in the 15-Minute City concept\cite{moreno_introducing_2021}. Stay events where the locations are Home or Work correspond directly to the MEs of the same name. MEs for stay events at 'Other' locations are assigned based on the type of activities performed during the event. For example, the activity `eating/drinking' corresponds to the Hospitality ME and `medical visit' to the Healthcare ME. Detailed relationships between activities and microenvironments are shown in Table~\ref{table1} (Supplementary Information, SI). The mapping of MEs to activities is a key step in this modelling framework because DAVE generates travel demands as people switch between different microenvironments. The above process creates a database of about 160,000 activity profiles (see the aggregated diurnal ME occupancy patterns in Fig.~\ref{sfig0}). For each simulation time step, citizens are assigned an activity profile probabilistically based on their age group, current time step, current ME, season, and day types (weekday, weekend, weekday without schools). In most cases, the citizen's profile is selected from the input profiles with an exact match of these markers. If an exact match is not found, the model relaxes the selection constraints following the order of time step, season, day type and ME.

\subsubsection*{Spatial attractors}
The activity profiles inform DAVE's temporal dimension of human behaviour while spatial attractors specify the likelihood of visiting different neighbourhoods (i.e., the 500~m simulation grid cells) in the study domain. Spatial attractors are used to rank the attractiveness of each ME in every 500~m grid cell for someone located in any other grid cell of the domain. The calculation is mostly based on the number of Point-of-Interests (POIs) or area related to the specific microenvironment in the destination grid cell and the distance between origin and destination \cite{capel-timms_dynamic_2020}. Residential population \cite{noauthor_population_2021}, working population \cite{noauthor_employed_2020}, schools \cite{noauthor_batiments_2025, deshogues_ecole_2006}, and land use data \cite{noauthor_swiss_2021-1} are available from official sources. POIs from OpenStreetMap \cite{noauthor_openstreetmap_2024} are extracted for facilities relevant to the MEs, such as restaurants, hospitals, and shops. The OSM tagging scheme is extensive and covers hundreds of categories that include the MEs considered here. POIs are linked to MEs through the relationships listed in Table~\ref{table2}. The total counts of POIs--ME pairs are then aggregated to the 500~m grid level. Note that there are a few MEs that are not based on POIs. For example, for the  Other Home ME, the residential population from the Federal Statistical Office \cite{noauthor_population_2021} is aggregated to the 500~m grid and used for calculating attractors (Fig. \ref{fig2}a). The spatial attractors of the Outdoor ME are calculated from land use statistics based on the land coverage nomenclature (NOLC04) at 100~m resolution \cite{noauthor_swiss_2021}. Only 500~m grid cells that have more than half of their area covered by categories of `Green spaces and recreation areas', `Alpine pastures', `Lake and rivers', or `Forests' are considered as attractive destinations for the `Outdoor' ME. The area of these land cover types in the grid cell is then used as a measure of attractiveness. Maps of the aggregated counts of POIs (or area for Outdoor ME) at 500~m resolution for each ME are shown in Fig.~\ref{sfig1}.

After aggregating the measures of attractiveness to the grid scale, spatial attractors are calculated by applying a reduction function based on the distance between origin and destination as follows \cite{hertwig_connecting_2025}:
\begin{eqnarray}
\label{eq:gravity}
\Gamma^{ME} = \frac{N_d^{ME}}{(\frac{D_{o,d}}{\Delta_N})^{\alpha}}
\end{eqnarray}

\noindent where $\Gamma^{ME}$ is the spatial attractor between origin ($o$) and destination ($d$) for the specific microenvironment,$N_{d}^{ME}$ is the number of POIs for that ME (or the number of 100 m\textsuperscript{2} green areas for Outdoor ME, or the residential population for Other Home ME) in the destination grid cell, $\Delta N$ is the grid scale of 500~m, and $D_{o,d}$ is the straight-line distance between origin and destination grid-cell centre points \cite{hertwig_connecting_2025}. The exponent $\alpha$ determines how strongly the attractiveness reduces with distance and it is currently set as 1.5 for all MEs (except Home and Work) \cite{schatzmann_spatial_2019, lenormand_systematic_2016}. Apart from this constraint, a catchment area is set for each ME to restrict the maximum travelled distance and reduce the computation time. There is no distance limit for the University and Hotel MEs; the limit is 35~km for Primary and Secondary School, and 20~km for the rest (all are larger than the 10~km used for London which is a more urbanized and compact region). The 35~km for schools is chosen to assure those living in remote rural areas also get assigned a school. For the other MEs, 20~km is deemed appropriate as it far exceeds the average Swiss daily travel distance for each trip purpose, i.e., 4.6~km for shopping and 12.9~km for leisure \cite{noauthor_swiss_2021}. This can be adjusted to allow/limit people travelling to farther/closer destinations. 

Deriving spatial attractors of the Work ME follows a different approach by using census data \cite{noauthor_employed_2020} (Fig.~\ref{sfig2}) that provide relations between population residence and workplace at commune level (2,202 communes existed in Switzerland in 2020). This dataset shows the number of people living in an origin commune ($o$) and working in a destination commune ($d$) for all possible combinations including within the same commune. Here, we use this information as the basis for calculating spatial attractors for the Work ME. First, the probability of working in all possible communes is calculated for each residence commune. Second, the working population in the destination commune is downscaled to 500~m grid level based on the ratio of non-residential building volume in each grid cell to the total non-residential building volume of the entire destination commune. Finally, the probability of working in a given destination grid cell is assumed to be the same across the grid cells in the residence commune. The spatial attractor for the Work ME uses these residence-work probabilities between grid cells, instead of Eq.~\ref{eq:gravity}, to represent more realistic commuting behaviours in the model. The above calculations for non-Work and Work MEs closely follows Hertwig et al. \cite{hertwig_connecting_2025} where more details can be found.

The spatial attractors set the probability weights of an individual visiting different grid cells within the spatial domain whenever they change MEs. This follows a weighted random choice process each time except for Work, Primary School, and Secondary School, as their locations are assigned only once at model initialisation, assuming people do not change their places of work or education during the simulation period. 

\subsubsection*{Population}
The Home ME of each modelled citizen is assigned at the start of the simulation based on residential population data (STATPOP) of the Federal Statistical Office \cite{noauthor_population_2021}. STATPOP gives population counts (total and stratified by age in 5-year bins) at 100~m resolution. The 100-m data are aggregated to 500~m to match the resolution of the DAVE simulation. Employment rates at commune level \cite{noauthor_economic_2021} are downscaled to 500~m grid level to set the percentage of the working adult population in each grid cell. The residential population is used as the initial condition of the simulation (i.e., all individuals are in their Home grid cell at the first time step). In the current Swiss simulation, we only separate 2 age groups, 18-65 (adults) and above 65 (seniors), while people under 18 years old are excluded due to the lack of activity profiles from TimeUse+. Although children are not included in our simulation, Primary and Secondary School MEs are kept for consistency with previous work \cite{hertwig_connecting_2025} and because adults can still visit these MEs, e.g., to pick up their children as part of their activity profiles. In the future, different cohorts can be defined according to the research objective. 

\subsubsection*{Transport database}
% overview
The simulation of citizens' movements between grid cells uses four transport modes (public transport, driving, cycling, and walking). A comprehensive travel database is generated from road network data \cite{noauthor_swisstlm3d_2024} and public transport timetables \cite{noauthor_general_nodate} following the MATSDA-roads v1.0 and MATSDA-metro v1.0 processing procedures \cite{hertwig_connecting_2025}. The travel database represents the transport network as a collection of connected nodes at grid-cell resolution, based on which MATSDA's pathfinder finds the optimal travel route between all grid-cell pairs for each travel mode using Dijkstra's algorithm, a well-known `shortest-path' method \cite{dijkstra_note_1959}. Route optimisation can consider different weights, e.g., travel duration (used in this study) or travel cost. In case of the former, the best routes from a starting grid cell to all other grid cells are determined by iteratively selecting the unvisited grid cell with the shortest travel time and updating the times to its neighbours. The process continues until all grid cells have been visited and the fastest routes are found. This is done offline prior to the main simulation so the mode-specific grid-to-grid fastest routes are pre-determined. The main process is described below while more details on the MATSDA version used can be found in \cite{hertwig_connecting_2025}. The recent release of MATSDA-roads v2.0 \cite{ma_matsda-roads_2025} introduces more detail by considering road-type nodes at sub-grid level, but was not available yet at the time of this study.

% private road types, modes, and speed 
Input data for private transport (driving, cycling, and walking) are retrieved from SwissTLM3D \cite{noauthor_swisstlm3d_2024} which contains information on the geometry and road types for every road in Switzerland. The dataset is processed with the MATSDA-roads v1.0 processor \cite{hertwig_connecting_2025} to represent realistic road connections between grid cells (i.e., a direct trip would not be possible for two adjacent grid cells without connected roads). The median speed is assigned to each road by subtracting 10~\( \mathrm{km\,h^{-1}} \) from the official speed limit given its road type \cite{noauthor_driving_nodate} and the typology (urban, peri-urban, or rural) of the communes \cite{noauthor_spatial_2020} in which the road is located primarily. For instance, 110~\( \mathrm{km\,h^{-1}} \) for highways, 90~\( \mathrm{km\,h^{-1}} \) for expressways, 60~\( \mathrm{km\,h^{-1}} \) for roads outside built-up areas, and 20~\( \mathrm{km\,h^{-1}} \) for local roads inside urban agglomerations. Similar methods have been used in Switzerland before \cite{loreti_local_2022}. The prescribed road speed reflects the general speed on the roads without considering diurnal variations (which can in fact be considered if temporal speed data are available; see \cite{ma_transport_2026}). This may overestimate car speed and its usage particularly during peak hours. Note that, in this study, the travel distance to an adjacent grid is assumed to be the diagonal distance of the 500~m grid (so 707~m) which is only an approximation of the realistic road length \cite{hertwig_multi-scale_2025}. This approximation may overestimate the distance and time for short walking trips and thus discourage citizens from choosing walking (this limitation is addressed in \cite{ma_transport_2026}). Cycling and walking are allowed on all road types except highways. Cycling speed is set at 20 \( \mathrm{km\,h^{-1}} \) on roads wider than 6~m or 15~\( \mathrm{km\,h^{-1}} \) otherwise and walking speed is set constant at 5~\( \mathrm{km\,h^{-1}} \) regardless of road type and location.  

% public transport
The main public transport modes considered in the study are train, bus, and metro, the latter of which only exists in the Lausanne metropolitan area. Their schedules are taken from the static General Transit Feed Specification (GTFS) for 2022 \cite{noauthor_general_nodate}. This GTFS file gives comprehensive information on public transport routes and schedules for all public transport stations in Switzerland. To match the simulation resolution, all stations of a specific mode in each 500~m grid cell are regarded as one station. The fastest connection between each origin-destination grid pair is determined from all the public transport connections between them. During the simulation, trips between grids for each mode follow the fastest connection by assuming people always choose the fastest option. In reality, the fastest connection (e.g., a direct train) is not always available and waiting time is often needed for public transport. Thus, to have more realistic travel times on public transportation, waiting times of 3, 2, and 1~min are added for trains, bus, and metro, respectively, at each stop during the trip. Temporal capacity of each public transport mode at 10-min resolution is included in the capacity dataset and input to DAVE \cite{hertwig_connecting_2025}. Therefore, people can only use the mode when it is available (for instance, there are no train departures in most grid cells after 1~am) and differences between weekday/weekend schedules of the same services can be captured. The model allows walking to reach public transport stops and transfers between public transport modes (e.g., bus to train).

% mode choice
A citizen's mode choice is determined in several steps. First, only the modes that are available for the route and still have capacity are considered (e.g., metro is only available in Lausanne area, no trains after 1~am, no cars at a place without road). Second, mode splits are set based on trip purposes (work, leisure, and shopping) using statistics of the Swiss Transport and Mobility Microcensus 2021 (MTMC) \cite{noauthor_swiss_2021-1}. For instance, the mode split for shopping trips is 42\% cars, 42\% foot, 6\% bikes, 8\% public transportation. The mode split for work is initiated differently for residents in Vaud and Geneva Cantons with the main difference being the higher public transport use in Geneva (43\% versus 31\% in Vaud) \cite{noauthor_swiss_2021}. However, a mode is discarded if its simulated travel time exceeds twice the average travel time reported in the MTMC for that trip purpose. A similar principle using twice the shortest-path duration as a feasibility filter has been applied in previous Swiss studies \cite{marra_determining_2020, marra_how_2023}. Third, walking mode is chosen if the destination is reachable within 10~min of walking, reflecting an average walking time \cite{noauthor_swiss_2021-1}. Finally, if no mode is selected through the preceding method, a mode is assigned probabilistically, with selection probabilities proportional to the inverse of each mode’s travel time. The commuting mode choice of each citizen is only computed once while the mode choices for other purposes are computed for each trip. The statistics on mode split from MTMC reflect the overall population behaviour which also reveals the effects of other socio-economic factors like car/bike ownership and public transport subscription that are not explicitly considered in our simulation. This method ensures that the overall mode split agrees with the general behaviour of the Swiss population.

\subsection*{Evaluation datasets}
To evaluate the simulation results, we use the mobility data from Swiss automatic road traffic counts (SARTC) \cite{noauthor_swiss_2022},  Swiss National Passenger Traffic Model (NPVM) \cite{noauthor_load_2025},  and MTMC \cite{noauthor_swiss_2021}. SARTC consists of a network of automatic counting stations on Switzerland's most critical roads where vehicles are detected and counted using induction loops or overhead laser sensors for both directions. Here we use the average weekday traffic values for July 2022, corresponding to our simulation period. NPVM is a macroscopic transport model managed by the Transport Modeling Office of the Federal Department. It uses a modular structure and about 8000 'traffic zones' to generate, distribute, and allocate the trips by different transport modes and networks in the country. The publicly-available output is for the base year 2017 and in the format of road traffic flows where we use the average weekday car values and adjust with the the measurements from SARTC. Finally, MTMC is the largest nationwide survey on mobility behaviour in Switzerland which is conducted every 5 years by the Federal Statistical Office (FSO) and the Federal Office for Spatial Development (ARE). We use the results from the most recent year (2021), consisting of about 56,000 respondents interviewed by telephone with online questionnaire. MTMC provides comprehensive travel behaviour statistics such as the ownership of vehicles/travel cards, travel time/distance/stages, travel routes, and travel modes. We evaluate our simulation with the temporal share of population on transport modes from MTMC, which is not directly related to the fixed mode splits that are part of the inputs. 

\section*{Results}
\subsection*{Dynamic population distribution}
\begin{figure}[!h]
\includegraphics[width=0.9\textwidth]{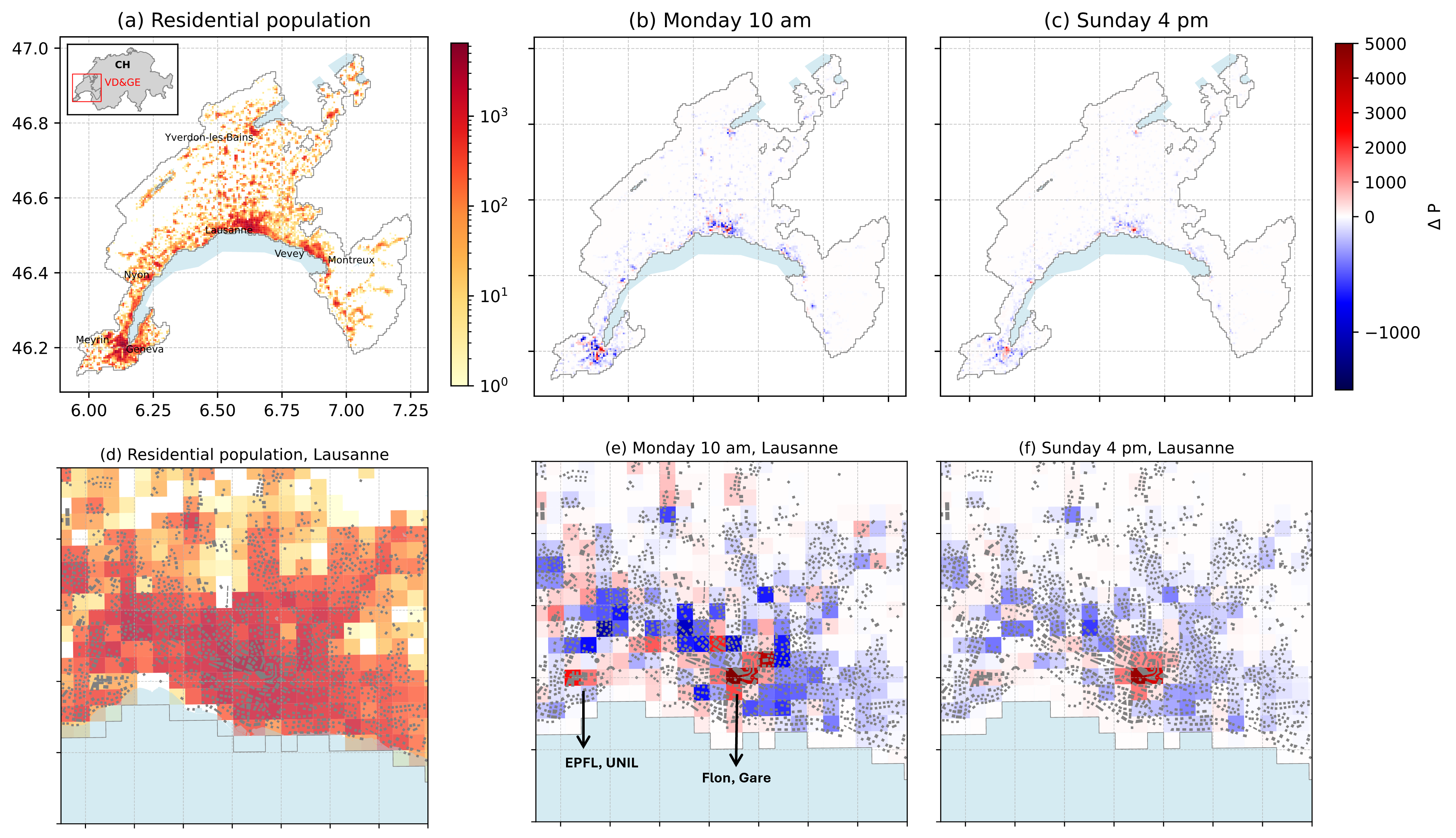} 
\caption{{\bf Adult population (18-65) distribution at different hours and days of the week.}
(a) Residential population ($P$) in Canton Vaud and Geneva (inset: VD\&GE) (b) Adult population difference ($\Delta P$) between Monday 10~am and residential population (c) Population difference between Sunday 4~pm and residential population. (d)–(f) show the same results as (a)–(c), but are zoomed in to the Lausanne region. Gray polygons in (d)-(f) are buildings.}
\label{fig2}
\end{figure}

% Total population 
Daily mobility reshapes the population distribution as people visit different neighbourhoods for various purposes. These dynamics are explored by comparing temporally-varying neighbourhood-level population to the spatial distribution of residential population (Fig.~\ref{fig2}). The residential population is much higher in the main urban agglomerations of Geneva, Lausanne, Yverdon, Nyon, and Montreux, compared to the rural areas in the Cantons. While most people are at their residence neighbourhood at night, people increasingly visit other neighbourhoods during daytime. During weekdays, the daytime population distribution is largely shifted to working places, particularly in the main urban centres of Geneva and Lausanne (Fig.~\ref{fig2}b). The most attractive areas are, for example, the central business districts near Lausanne Flon and Gare districts and the university campus of EPFL and UNIL on the west side of the city (Fig.~\ref{fig2}e). This result is consistent with the spatial attractors of the Work ME input to the model (Fig.~\ref{sfig2}). During weekends, the same city districts do not attract as many people but the population in urban areas still increases because of the high density of leisure and hospitality locations compared to the (rural) surroundings (Fig.~\ref{fig2}c,f). This explains why some urban regions between Lausanne and Geneva, like Nyon and Morges, show high incoming populations during the weekend even if the same increase is not necessarily evident during weekdays. Therefore, different weekend--weekday mobility patterns emerge depending on the most prominent microenvironments of each neighbourhood.

\begin{figure}[!h]
\includegraphics[width=0.9\textwidth]{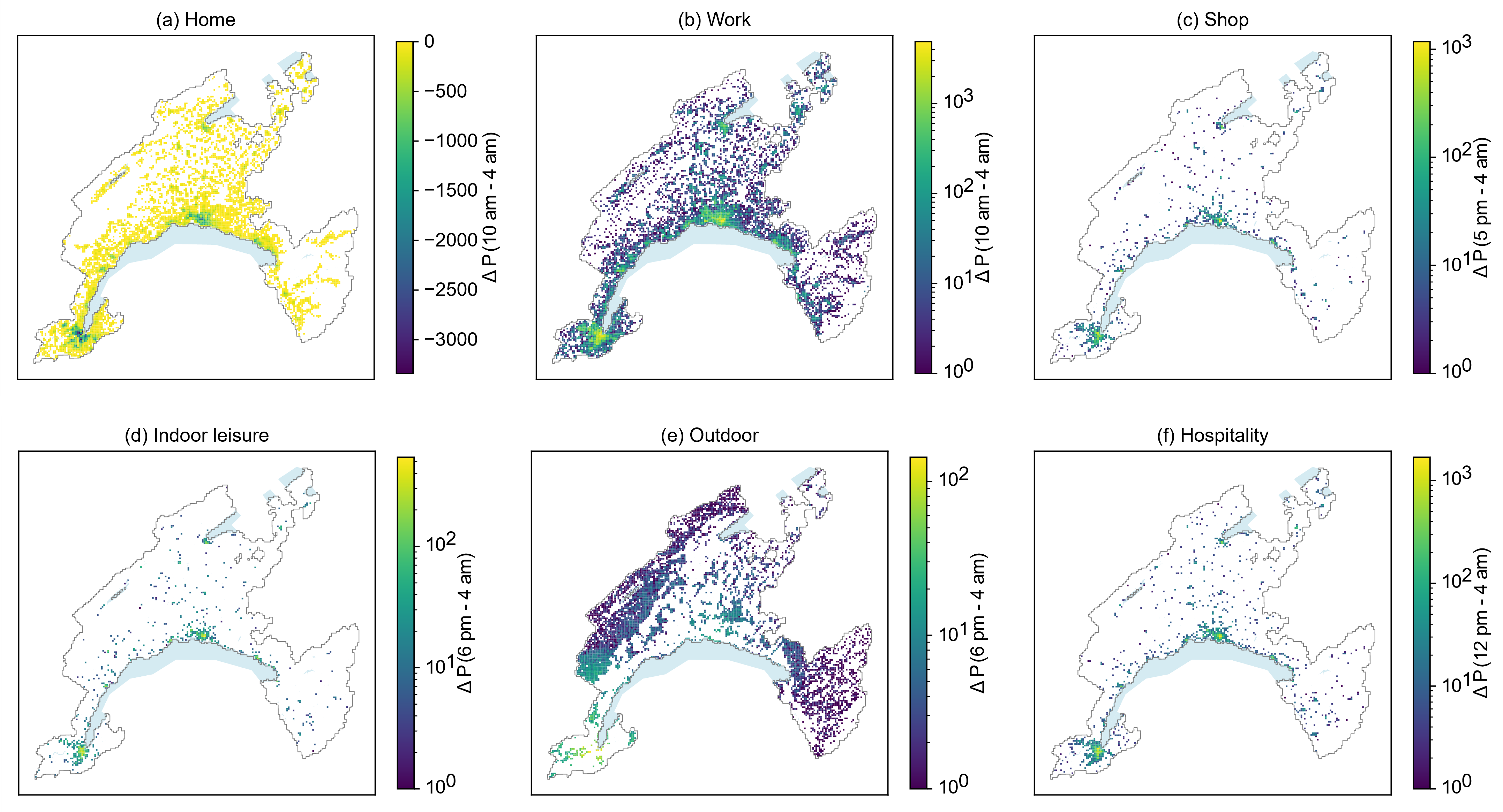} 
\caption{{\bf Population changes $\Delta P$ in different microenvironments.}
Population difference ($\Delta P$) between peak or low hours and 4~am (color bar) for adults on a weekday for six MEs: (a) Home (low: 10~am), (b) Work  (peak: 10~am), (c) Shop (peak: 5~pm), (d) Indoor Leisure (peak: 6~pm), (e) Outdoor (peak: 6~pm), and (f) Hospitality (peak: 12~pm).}
\label{fig3}
\end{figure}

% Population in specific ME
Within each neighbourhood, DAVE tracks which microenvironments citizens occupy based on their activities. Fig.~\ref{fig3} shows adult population numbers in the dominant MEs in the domain. For each ME, we show the difference in population numbers between its peak hours (except for Home ME) and early morning (4~am, population distribution similar to residential population) to allow for easier comparison. Home and Work are the two MEs that experience the largest adult population changes on a weekday (Fig.~\ref{fig3}a,b) so the transitions between them largely shape the weekday adult population dynamics. The other MEs attract people during certain times of the day. For example, Hospitality has higher population at lunch and dinner time; the Indoor Leisure, Shop, and Outdoor MEs attract population on weekday evening as people leave the Work ME. The population distributions in different MEs (Fig.~\ref{fig3}) are closely related to the distribution of POIs of the ME across neighbourhoods (Fig.~\ref{sfig1},\ref{sfig2}). In other words, when people want to visit a certain ME, they are attracted to the neighbourhoods that are rich in that service or activity, but this is modulated by origin-destination distances and travel modes. One example is the high attractiveness of Outdoor ME neighbourhoods, with parks scattered in Geneva and the forest in the north of Lausanne attracting more people in the evening when there is a tendency to visit outdoor spaces. Another example is the dense Indoor Leisure ME in the Geneva and Lausanne urban centres that attracts a high influx of visitors in the evening.

% Population flows
\subsection*{Population flows}
\begin{figure}[!h]
\includegraphics[width=0.9\textwidth]{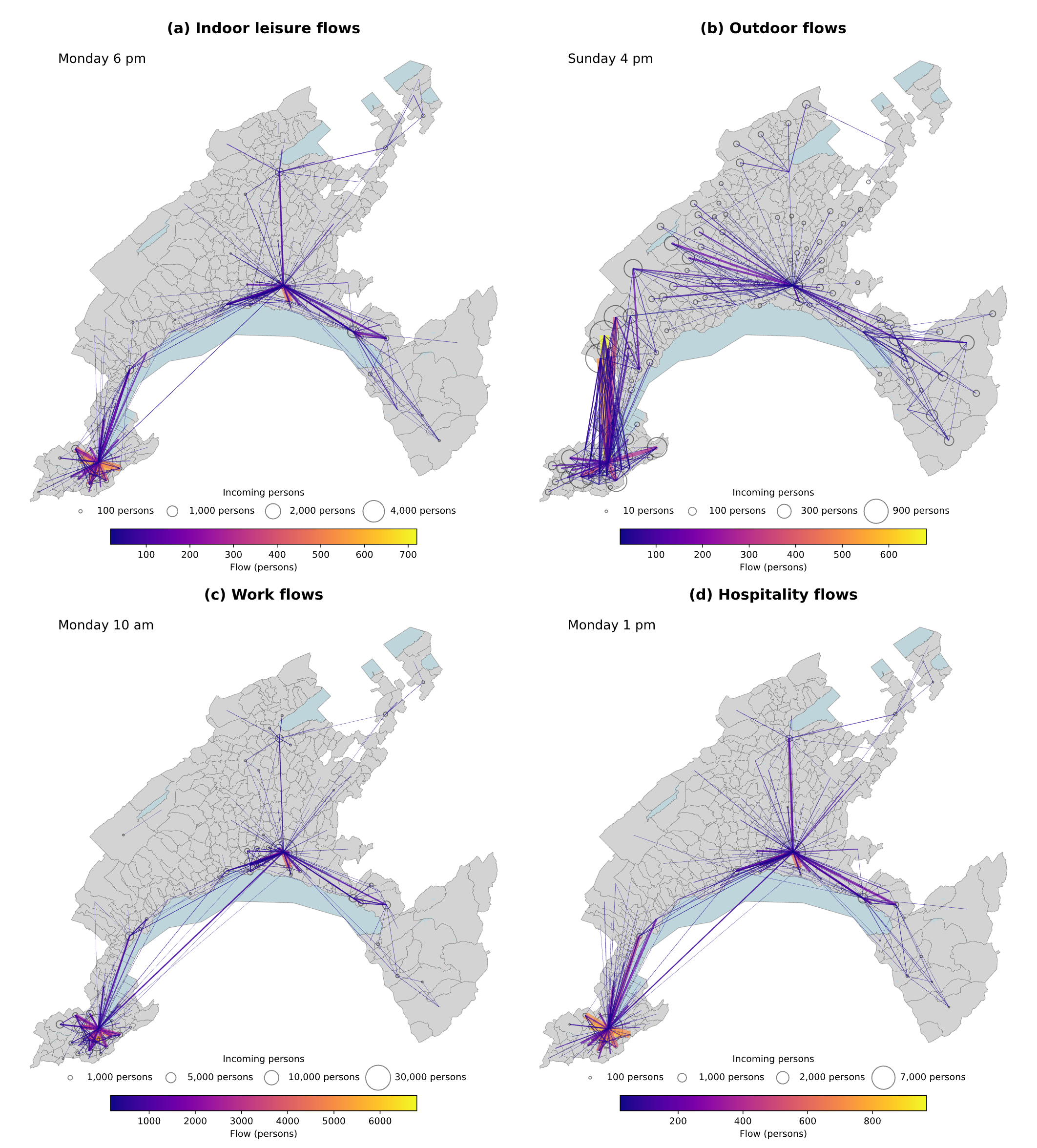} 
\caption{{\bf Population flows for accessing different microenvironments at commune resolution.}
Population flows between residence and destination for accessing (a) Indoor Leisure, (b) Outdoor, (c) Work, and (d) Hospitality MEs on different days and times, with intensity of the travel flows (line thickness, colours) and the number of incoming populations (circle size) shown at the geographical centres of the 350 communes of the study region. The flows are shown as straight lines between the origins and destinations instead of the full travel routes.}
\label{fig4}
\end{figure}

The simulated mobility flows to access different MEs highlight the emergence of unique patterns depending on the spatial distribution of different socio-economic settings within the study region, demonstrated in Fig.~\ref{fig4} for four different MEs. Flows to the Indoor Leisure and Hospitality MEs mostly converge to urban centres particularly Geneva, followed by Lausanne and Yverdon, due to the high density of associated POIs. Although flows to the Work ME are higher in those cities, some smaller urban areas (e.g., on the west of Lausanne and outside Geneva) attract people too, showing a larger spatial spread of trip destinations. By contrast, flows to the Outdoor ME are mostly directed to the outskirts of cities and much more decentralized, meaning people have more options for choosing where to do outdoor activities. Population counts for different MEs follow distinct spatial distributions with the Indoor Leisure ME being the most spatially variable and Outdoor ME the most similar among the neighbourhoods evaluated (Fig.~\ref{sfig3}). 

\subsection*{Temporal population variation}
\begin{figure}[!h]
\includegraphics[width=0.9\textwidth]{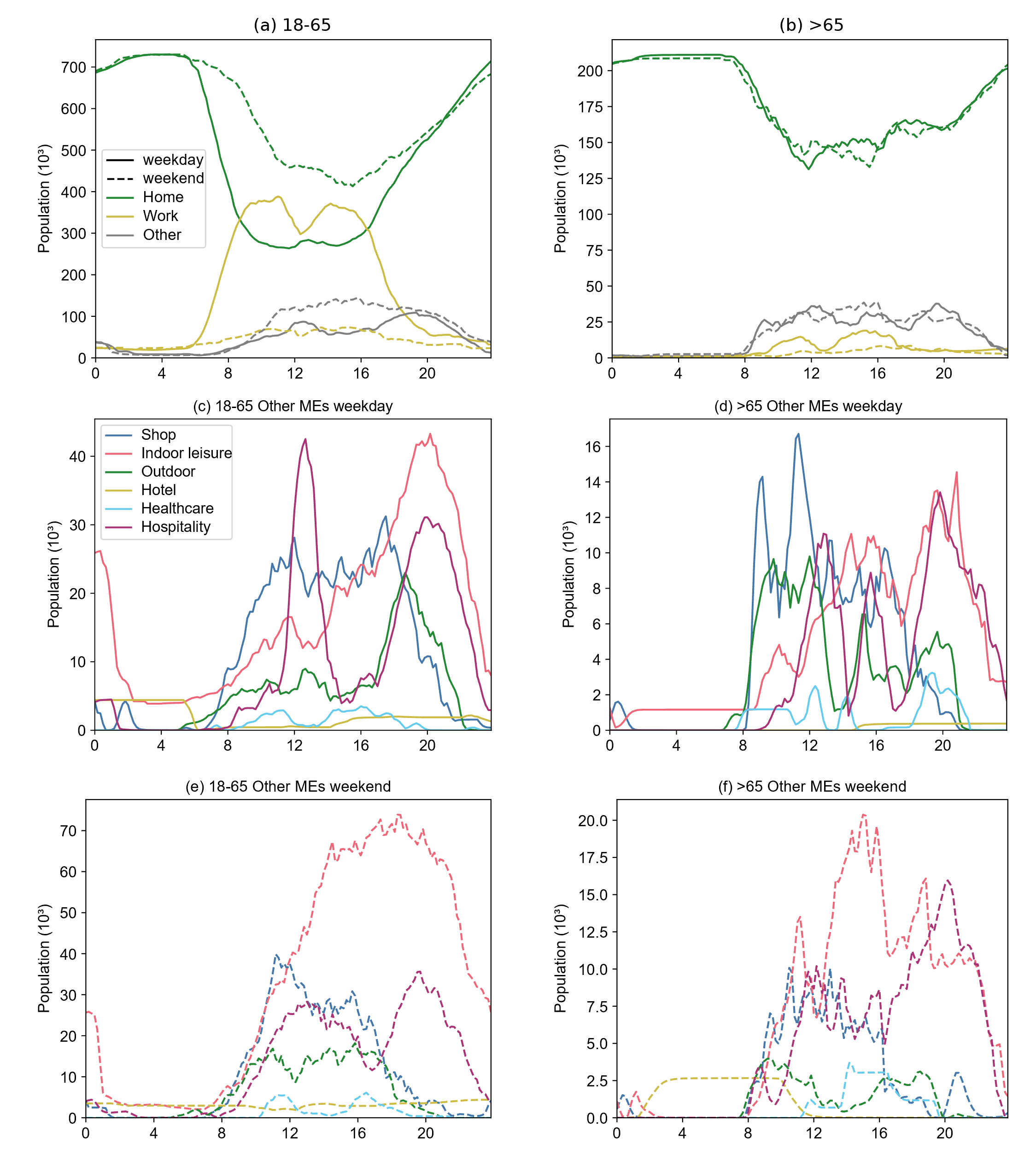} 
\caption{{\bf Diurnal ME occupancy profiles for different age group totals.}
Population of both Cantons in different MEs at 10-min resolution for (a,c,e) adults (age 18-65) and (b,d,f) seniors ($>$65). (a,b) profiles for `Home', `Work' and combined remaining MEs (`Other') for both day types; details of remaining MEs for (c,d) weekday and (e,f) weekend days.} 
\label{fig5}
\end{figure}

% change of population in ME
Citizens of the region simulated by DAVE visit various MEs throughout a day and behave differently depending on age, time and type of day (i.e., weekday or weekend). Figure~\ref{fig5} shows diurnal ME occupancy profiles derived over all weekday/weekend days of the simulation across the domain and stratified by population age (adults: Fig.~\ref{fig5}a,c,e; seniors: Fig.~\ref{fig5}b,d,f). Daily profiles for a weekday for adults largely depend on working hours, which typically are 8~am to 5~pm in Switzerland. A sharp reduction of population at work occurs around noon in response to the norms for timing of lunch (Fig.~\ref{fig5}a) with a corresponding sharp peak in the Hospitality ME (Fig.~\ref{fig5}c). In the evening, adults leave the Work ME and travel towards Home or other MEs like Shop, Outside, Hospitality, or Indoor Leisure (Fig.~\ref{fig5}c). The detailed profiles in Fig.~\ref{fig5}c show that the Indoor Leisure ME attracts most working age adults during the evening while the Shop ME has a rather stable population throughout the day. The Hospitality ME has a sharper increase in adults at noon compared to the wider secondary peak in the evening, showing that most adults have lunch at the same time (temporally constrained by work hours) while dinner time is more flexible.

% weekend
Adults' occupancy profiles show strong differences between weekday and weekend in response to the weekday-dominated work patterns of most adults (Fig.~\ref{fig5}a). In general, more than half of the adults stay at home during weekends. For those leaving the Home ME, this is delayed by 1-2~hours compared to weekday. The most visited ME during weekend mornings is Shop for weekly grocery shopping (Fig.~\ref{fig5}e). Around noon, the population in the Hospitality ME increases but not as sharply as the weekday peak. In the afternoon, more adults than on a weekday are  conducting activities in an Indoor Leisure setting. The Outdoor ME attracts a relatively constant number of people between 9~am and 6~pm on weekends, while there is a clear evening peak (after work) on weekdays. In general during the weekend, the diurnal occupancy profiles are less peaked because of more flexible schedules.

% senior
Different from the working-age adults, people older than 65 in Switzerland are mostly retired and only few still work (Fig.~\ref{fig5}b). More than half of the senior age group stays at home independent of day type. On a weekday, those leaving the Home ME mostly go to the Shop or Outside ME in the morning, or Indoor Leisure ME in the afternoon (Fig.~\ref{fig5}d), similar to adults' behaviour on weekends. Visits of the Hospitality MEs by seniors have distinct peaks around noon, 4~pm and 7~pm compared to two peaks for the working age adults. As they are less defined by work-pattern constraints, the senior cohort activity profiles are quite similar on both day types. One difference is that fewer seniors visit the Shop ME on weekend mornings, possibly to avoid adult shoppers at this time. Instead, seniors are more likely to frequent the Indoor Leisure or Hospitality MEs during that time (earlier compared to adults) as larger variety of cultural activities occur then compared to weekday mornings. All the simulated temporal profiles in Fig.~\ref{fig5} are consistent with the input TimeUse+ profiles (Fig~\ref{sfig0}). 

\subsection*{Transport modes and evaluation}
\begin{figure}[!h]
\includegraphics[width=0.9\textwidth]{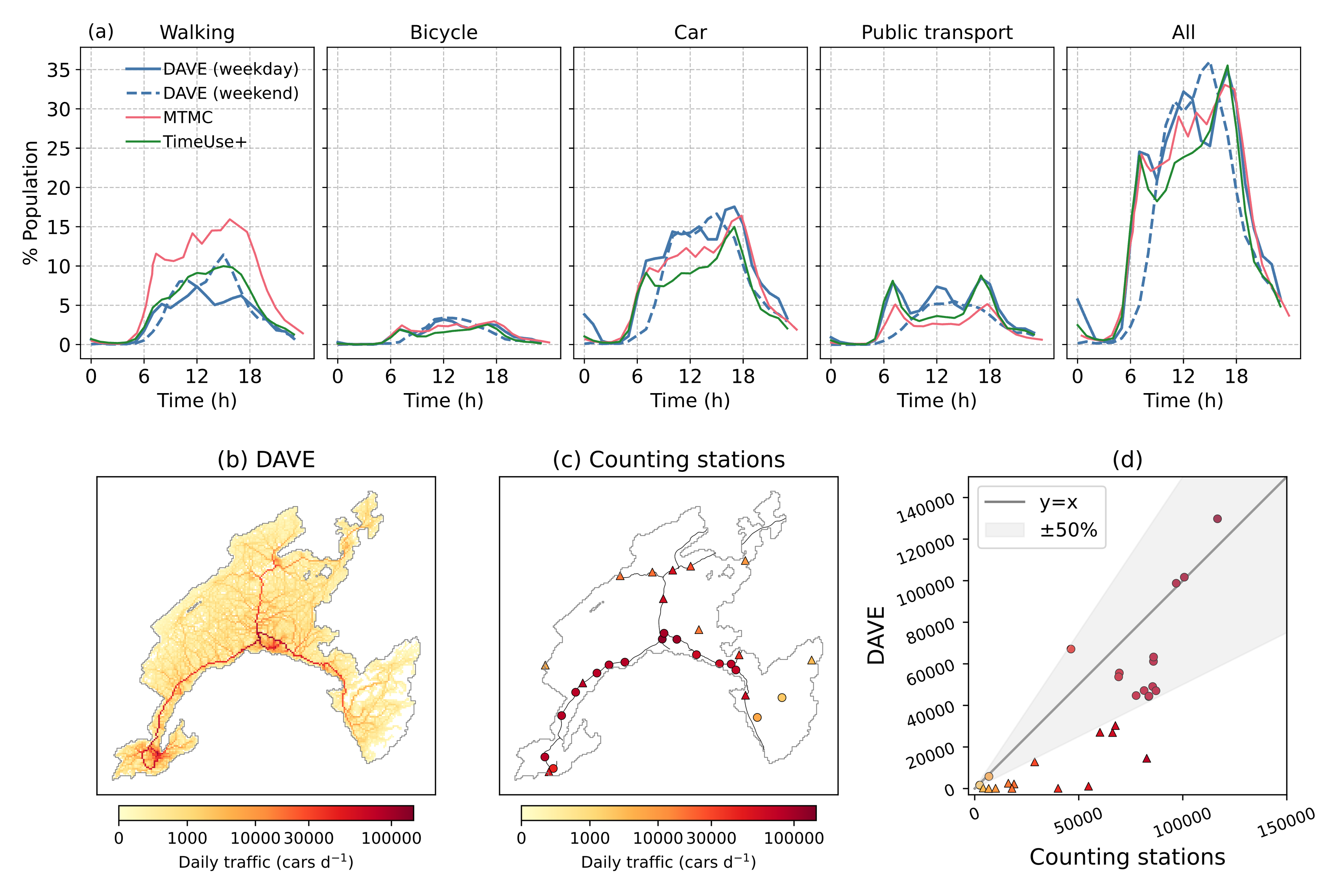} 
\caption{{\bf Temporal transport mode splits and spatial distribution of traffic flows.}
(a) Percentage of population on the move on different transport modes from DAVE (by day type) and observed MTMC and TimeUse+ reference data (all day types), (b) weekday car flows at 500~m resolution from DAVE, (c) weekday daily average measured car flows at 28 automatic counting stations \cite{noauthor_swiss_2022} (triangles: where DAVE deviates more than $\pm$50\%) mostly along highways (gray lines), and (d) comparison between DAVE modelled and observed car counts with $\pm$50\% (shading).}
\label{fig6}
\end{figure}

% temporal weekday
In addition to the population in different microenvironments, DAVE also provides occupancy profiles of different transport modes. Figure~\ref{fig6}a shows the simulated temporal profile of the population by transport mode in comparison to statistics from MTMC and TimeUse+. The overall percentage of people on the move in DAVE agrees well with observation-based statistics, with the highest peak ($\sim$32 \%) in the evening, preceded by a broader noon peak, and narrower peak during morning rush hour. In each transport mode except walking, DAVE has clear maxima during the peak morning and evening hours associated with travel between Work and Home MEs. Cars are the main mode, followed by walking, public transport, and bicycles. The walking population does not have distinct peaks but a constant percentage of about 7 \% from 8~am to 6~pm because many trips involve a short walk (e.g., from the residence to the nearest train station or bus stop). However, the modelled walking proportions are lower than the MTMC data ($\sim$15 \%) because except for very short trips (within 10 minutes) or the trips for work, leisure, and shopping, citizens prefer to choose the fastest mode available (see Methods). In reality, people may opt for walking for other benefits such as health and cost savings even though it is slower. Another major difference between DAVE and the other datasets is the higher noon peaks, likely due to the lack of temporally-varying mode choice information in the input data (see Discussion). 

% temporal weekend
During the weekend, the distinct morning and evening peaks caused by commuting are replaced by a broad plateau in the afternoon for car and public transport modes. This aligns with the timing of leisure/shopping ME occupancies (Fig.~\ref{fig5}) on weekends, as people's schedules become more flexible than on weekdays. Similar to the weekdays, most trips are car-based during the weekend but more concentrated between the hours of 9~am to 3~pm. However, the percentage of people walking increases on the weekend, reflecting the higher prescribed walking probabilities for leisure and shopping activities (the primary trip purposes on the weekend) than for work commuting (the primary purpose on weekdays). 

% spatial 
In terms of spatial distribution, the modelled daily car traffic is highly concentrated on the highways that link major cities (e.g., Geneva-Lausanne and Montreux-Lausanne) as they usually provide a faster connection than local roads (Fig.~\ref{fig6}b). The rural and mountainous areas in the east and west have much less traffic due to lower population densities. It is possible to compare DAVE results with the average weekday traffic measured by the SARTC \cite{noauthor_swiss_2022}. Simulation results can be most robustly compared with the reference data near the centre of the model domain (Fig~\ref{fig6}c,d). As the simulation does not consider vehicles travelling to and from outside the domain, the roads at the boundaries (triangles, Fig.~\ref{fig6}c,d) are less frequented compared to the observations. We also compare the results with the simulations from NPVM \cite{noauthor_load_2025} in Fig.~\ref{sfig4}. This comparison shows that the main mobility patterns on the inter-city highways by DAVE are consistent with NPVM simulations, but underestimations occur near the domain borders. The average daily driving and total travel time from DAVE (38~min and 82~min) are close to 30~min and 73~min observed in MTMC (Fig.~\ref{sfig6}). The slightly longer time is due to the higher proportion of younger groups (18-40) in TimeUse+ compared to MTMC \cite{winkler_timeuse_2024}. 
 
\subsection*{Spatial variation in mobility behaviour}
\begin{figure}[!h]
\includegraphics[width=1\textwidth]{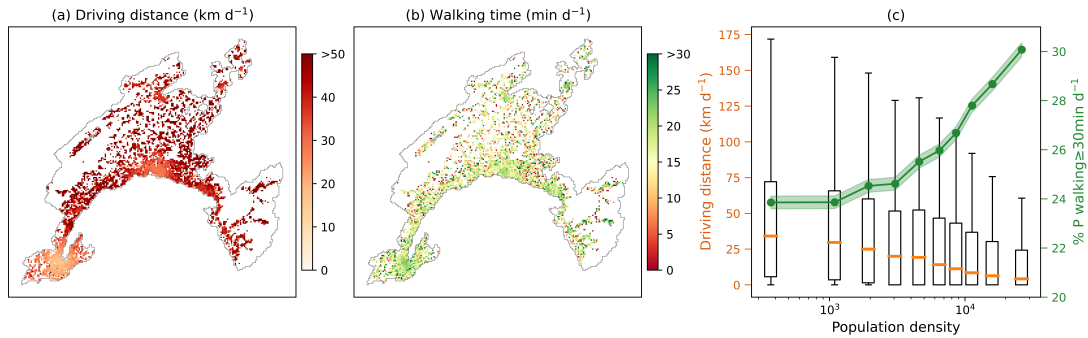} 
\caption{{\bf DAVE modelled average mobility behaviour in each grid-cell neighbourhood.}
(a) Average daily driving distance per person in each neighbourhood, (b) average daily walking time in each neighbourhood, (c) mobility metrics by residential population density: daily driving distances (box plots showing the median and IQR) and percentage of population walking $\geq 30$ minutes per day (shading: 95\% CI).}
\label{fig7}
\end{figure}

Traffic flows depend on mobility mode accessibility and mode choice. DAVE simulations reveal different transport preferences across neighbourhoods that are linked to public transport availability and connectivity, and to the spatial distribution of essential socio-economic settings. In urban areas like Lausanne and Geneva where good public transport services and diverse infrastructure exist, average driving distances are much shorter than in rural areas (Fig.~\ref{fig7}a). As residents of rural neighbourhoods have to travel farther to access different places of activity, and are more dependent on the use of private cars, their driving distances are longer on average. Urban dwellers also walk longer than people living in rural areas  (Fig.~\ref{fig7}b) because they can more easily reach their desired destinations by walking or a combination of walking and public transport. In rural neighbourhoods, most people walk less than the 30 minutes per day, which is the  recommanded threshold for physical activity duration \cite{noauthor_who_2020}. On average, people living in higher population density neighbourhoods drive shorter distances and walk more than people living in lower-density areas (Fig.~\ref{fig7}c). This trend aligns with the MTMC data \cite{noauthor_swiss_2021} which show the Swiss urban population drives 10.1~km less and walks 5.7~min more per day than the rural population. Thus, DAVE captures neighbourhood mobility preferences while simultaneously reflecting important urban sustainability and health indicators (see Discussion).

\section*{Discussion}
In this study we applied the state-of-the-art agent-based model DAVE to simulate population dynamics in Cantons Vaud and Geneva in Switzerland at 500-m resolution, the first inter-city modelling of DAVE. Simulation results reconstruct the temporally varying distribution of the population at 10~min resolution in different neighbourhoods and microenvironments (ME) as modulated by human activity schedules, neighbourhood facilities (urban function), and transport connectivity. As such, the model results provide the foundations for a variety of future research applications, from activity related emissions (e.g., anthropogenic heat) to indoor/outdoor exposure to environmental stresses like heat or air pollution.

%\subsection*{Model strengths and limitations}
The activity-based nature of DAVE enables modelling of the behaviour of  different socio-demographic population groups according to their typical daily activities. The behaviour of different age groups is clearly distinguished in the diurnal sequence of activity spaces occupied (Fig.~\ref{fig5}). Although in this work we only separate the population into two age groups (adults and seniors), it is possible to account for other socio-demographic characteristics such as income, employment status or car ownership. This opens up the opportunity to address various research questions but is constrained by input data availability. The behaviour on different day types is also distinguished, providing more detailed insights into population dynamics (Figs.~\ref{fig1},\ref{fig5},\ref{fig6}). Another strength of DAVE is modelling population distribution and travel flows in/to different MEs, affected by the intra- and inter-city distribution of ME properties (Fig.~\ref{fig3},\ref{fig4}). The modelling of people in different microenvironments of activity at neighbourhood level (Fig.~\ref{fig3}) offers more detail than other common mobility-tracking methods such as call detail records, travel card records, or social media tags \cite{tjaden_measuring_2021}. This detailed information can improve our understanding of building energy use and exposure to mobility-modulated indoor and outdoor heat and air pollution \cite{lin_daily_2025, liu_dwelling_2024, huang_economic_2023}. Furthermore, modelling travel flows for different purposes (Fig.~\ref{fig4}) reveals the travel demands beyond the city boundaries and how ME properties in different regions complements each other. This provides insights for transport-oriented urban design through, e.g., re-balancing urban function among different neighbourhoods, increasing mixed land uses, and coordinating inter-city development plans. 

DAVE captures macroscopic patterns on mobility behaviours from simulating individual's travel decisions with limited macroscopic input parameters (e.g, public transport schedules and mode choice statistics, see Materials and Methods). As Fig.~\ref{fig7} shows, the spatially-varying mobility preferences emerge from DAVE simulations without prior input on neighbourhood-specific transport mode splits (i.e., the commuting mode split is only initialized differently between Vaud and Geneva Cantons but not finer spatial units). Therefore, the variation is caused by the inter-region distribution of the transport network, population age groups, amenities and places of work. Aggregating from individual mobility trips, key sustainability and health indicators such as the daily driving distance and walking time \cite{noauthor_transport_2022, lee_importance_2008, noauthor_who_2020} for each neighbourhood can be calculated. The driving and walking behaviour represents not just a neighbourhood's accessibility to essential social functions, but also reflects its car dependence, traffic energy demand, and potential health and social benefits from walking \cite{noauthor_who_2020,sonta_rethinking_2023}, which are critical to designing sustainable and liveable cities. The increasing trend of walking time with population density aligns with previous observations on global cities \cite{cerin_determining_2022}, showing that denser urban environments may encourage more physical activities (walking) to meet the recommanded health criteria \cite{noauthor_who_2020}. This enables further applications of DAVE to guide greener and healthier urban designs that are related to some widely-discussed urban development concepts like 15-Minute City \cite{moreno_introducing_2021} and Superblock model \cite{nieuwenhuijsen_superblock_2024}. The modular structure of DAVE also allows for testing different scenarios by altering model input, e.g., ageing and migration of population in different neighbourhoods, construction of new public transport lines (e.g., Jura–Léman–Salève rail lines) or development of residential/commercial neighbourhoods, all of which are currently underway in Canton Vaud and Geneva.  

Despite these advantages and potential, the current study has some limitations. For behaviour modelling with SHAPE, our simulation only considers people older than 18 as TimeUse+ does not include younger age groups. This issue can be addressed if time use data of younger age groups become available (e.g., \cite{hertwig_connecting_2025, capel-timms_dynamic_2020}. We also separate behaviour into day type groups of weekday and weekend, but different behavioural patterns exist on different days within these categories. For example, adults in Vaud and Geneva Cantons with primary-and secondary-school children are more likely to work from home on Wednesdays as the school time is half day only \cite{noauthor_official_nodate, noauthor_school_nodate}. Also, fewer people go to Shop and Hospitality MEs on Sundays compared to Saturdays as most shops and restaurants are closed. In general, these details can be accounted for by further stratifying the input activity profiles into days of the week, but this impacts the overall sample sizes for each age group on each day type and hence the statistical representativeness of results (e.g., the total senior age group in TimeUse+ is already small with 115 respondents, explaining the fluctuations observed in ME profiles in Fig.~\ref{fig5}). Given our input mobility data  \cite{noauthor_employed_2020, noauthor_swiss_2021}, the transport mode split mostly depends on trip purpose (separated by working, leisure, or shopping) without considering the time of day. For example, leisure trips (including going to restaurants) have a 34\% probability of driving and 13\% of using public transport. While this assumption may hold overall, fewer people drive or take public transport during their lunch break on a weekday, therefore overestimating cars and public transport use around noon while underestimating walking (Fig.~\ref{fig6}a). Other constant parameters in our simulation (for computational efficiency) include the exponent $\alpha$ in Eq.~\ref{eq:gravity} and the catchment radii for each ME. These choices result in similar travel distances per trip for different age groups, at different hours of the day, and between weekday and weekend. In reality, younger people travel longer distances; evening leisure trips are longer than noon lunch trips; and people travel longer during the weekend than weekday \cite{noauthor_swiss_2021}. Finally, the gravitational mobility law in Eq.~\ref{eq:gravity} may not be suitable for all MEs. For example, although there are many hotels in Geneva, the people living in the city most likely visit hotels further away for holidays. Transboundary traffic flows are not considered in this study (Fig.~\ref{fig6}) but cross-border commuters in fact account for 20\% of employees in central Geneva and even up to 50\% in some municipalities on the border \cite{noauthor_cross-border_nodate}. Cross-border trips for leisure and shopping are also common but not considered \cite{leimgruber_tourism_2021, burstein_cross-border_2024}. This limitation can be addressed by expanding the simulation domain or by statistically accounting for national and international visitors, including those at locations such as Geneva Airport. Besides these limitations used in our simulation, the DAVE model is currently under further development to improve many of its aspects e.g., refining the transport module to higher spatial resolution (sub-grid scale road level) and stronger coupling between the building energy and land-surface models  \cite{ma_transport_2026,liu_vertically_2026,hertwig_dave_2025}). 

%\subsection*{Implications and future study}
This study demonstrated the transferability of DAVE to other regions (London to Switzerland) and scales (one city to multiple cities). The main challenges include identifying and processing suitable local datasets (both for inputs and evaluation) that represent region-specific contexts (e.g., ME classes of interest, mobility statistics) and research needs. For inter-city modelling, an additional challenge may be the merging and harmonisation of data from different sources (i.e., Vaud and Geneva Cantons here) to consistent representation and format. This study focused in detail on the transport and behaviour modules of DAVE to test and evaluate the model and provide the basis for future human exposure studies that account for activity dependency. The building energy and land surface modules are not exploited here. Future research will focus on implementing the full DAVE model to describe the complex multi-sector interactions that characterise urban systems, for example, to assess population exposure to heat and pollution in indoor and outdoor environments considering urban microclimate and mobility. Currently, most heat exposure studies assume exposure to outdoor temperature whereas in reality people spend most of their time in indoor setting \cite{uejio_summer_2016, hondula_novel_2021}. Thanks to the integration of the STEBBS and SUEWS modules, DAVE can simulate both outdoor and building indoor temperatures thus allowing for more accurate heat exposure assessments \cite{liu_vertically_2026, hertwig_modelling_2024}.

\section*{Conclusion}
The interactions between humans and the form and functions of the built environment are generally oversimplified in top-down urban modelling. Few bottom-up or agent-based models exist (e.g., \cite{capel-timms_dynamic_2020, wang_cesar_2018, w_axhausen_multi-agent_2016}) but they usually focus on specific urban functions or neglect inter-city and urban-rural interactions, thus not fully reflecting the complexity of existing inter-connected urban systems. To start tackling these challenges, we applied the multi-facet agent-based model DAVE to simulate population activities and mobility in Canton Vaud and Geneva in Switzerland; a region of tightly-knitted urban agglomerations, rural areas, and alpine environments. The results show detailed age-specific population dynamics across neighbourhoods and microenvironments of activity using different transport modes. The weekday mobility of working-age people is largely shaped by morning commuting trips, noon lunch breaks, and after-work trips to leisure activities and people's residences. On the weekend, the adult population spends more time at home and in indoor leisure environments and the temporal occupancy profiles are smoother without morning and evening peaks. The behaviour and mobility patterns of the retirement-age population are more similar between weekdays and weekends with only small differences in the timing of activities in different microenvironments. The population variation in different neighbourhoods is modulated by their main social functions and residents' daily habits. Due to the uneven spatial distribution of points-of-interest, the flows to indoor leisure, hospitality, and commercial/shopping settings are much more concentrated in highly urbanised neighbourhoods compared to the more decentralised flows to workplace and outdoor environments. With its detailed consideration of the coupling between behaviour, mobility, and spatial environments, this study contributes to the discourse on data-driven urban planning and the agent-based modelling presented here. Furthermore, this work lays the foundations to address a variety of challenges in the transportation, energy, and urban development sectors. 

\section*{Acknowledgments}
GSL, GM, and MLLM acknowledge support from the SNSF Weave/Lead Agency funding scheme (grant number 213995). Funding by the European Research Council (ERC) under the European Union's Horizon 2020 research and innovation programme (ERC-SyG no. 855005; project \textit{urbisphere}; DH/MM/TM/SG) and by the UKRI Natural Environment Research Council (NERC) projects ASSURE (NE/W002965/1; DH/MM/STS/SG) and APEx (NE/T001887/2; DH/STS/SG) are gratefully acknowledged. We thank Prof. Kay W. Axhausen and Daniel Heimgartner at ETH Zurich for providing the TimeUse+ dataset. We thank Vaud Cantonal Administration and Territorial information system in Geneva (SITG) for providing data on the buildings in their cantons. 

% \nolinenumbers

% Either type in your references using
% Guoshiuan added these 2 lines - have to be deleted later on
% \bibliographystyle{plos2015} 
\bibliography{references}
\label{LastMainPage}
% \begin{thebibliography}{}
% Text
% \end{thebibliography}
%
% or
%
% Compile your BiBTeX database using our plos2015.bst
% style file and paste the contents of your .bbl file
% here. See http://journals.plos.org/plosone/s/latex for 
% \begin{thebibliography}{10}

% \bibitem{bib1}
% Conant GC, Wolfe KH.
% \newblock {{T}urning a hobby into a job: how duplicated genes find new
%   functions}.
% \newblock Nat Rev Genet. 2008 Dec;9(12):938--950.

% \end{thebibliography}

\clearpage
% \pagenumbering{gobble}
\pagestyle{empty}
\section*{Supplementary Information (SI)}
% Place tables after the first paragraph in which they are cited.
\renewcommand{\thetable}{SI~\arabic{table}}
\renewcommand{\thefigure}{SI~\arabic{figure}}

\begin{table}[htbp]
\caption{
{\bf Relationship table between TimeUse+ \cite{winkler_timeuse_2024} variables and microenvironments in DAVE}}
\centering
\begin{tabular}{|p{4cm}|p{4cm}|p{4cm}|}
\hline
TimeUse+ "event\_name\_imputed"& TimeUse+ "activities" & DAVE microenvironments\\ \hline
HOME & all& Home\\ \hline
Workplace& all& Work\\ \hline
Other& Working& Work\\ \hline
Other& Sleeping& Hotel if there is expenditure; Other Home otherwise\\ \hline
Other& Restaurant& Hospitality\\ \hline
Other& Hobby / Leisure& Indoor Leisure\\ \hline
Other& Errands, Shopping, or Package pick up / drop off& Shop\\ \hline
Other & Exercising or Walking the dog& Outdoor\\ \hline
Other& Studying& University\\ \hline
Other& Medical visit& Healthcare\\ \hline
Other& Eating / cooking, Resting, Self-care& Other Home\\ \hline
Other& Caretaking& Healthcare or Other Home \\ \hline
 Other& Socializing& Indoor Leisure, Other Home, Outdoor\\\hline
 Other& Drop off / pick up& Primary or Secondary School depending on ages of the trackers' kids. Random choice of all MEs if no kids\\\hline
\end{tabular}
\begin{flushleft} 
\end{flushleft}
\label{table1}
\end{table}
\clearpage
\setlength{\tabcolsep}{4pt}
\setlength{\LTleft}{0pt}
\setlength{\LTright}{0pt}
\begin{longtable}{|p{1.53cm}|p{1.5cm}|p{6cm}|p{4cm}|p{2.0cm}|}
\caption{{\bf Sources to calculate the spatial attractors for each microenvironment\label{table2}}} \\
\hline
\textbf{Microenv.} & \textbf{Sources} & \textbf{OSM tags} & \textbf{Specific dataset} & \textbf{Catchment radius (km)} \\ \hline
\endfirsthead
\hline
\textbf{Microenv.} & \textbf{Sources} & \textbf{OSM tags} & \textbf{Specific dataset} & \textbf{Catchment radius (km)} \\ \hline
\endhead

\endfoot
\endlastfoot

Home & FSO & NA & Residential population & NA \\ \hline
Work & FSO & NA & Commune-level residence-work pairs & NA \\ \hline
Indoor Leisure & OSM & theatre, cinema, sports\_centre, swimming\_pool, 
nightclub, ice\_rink, attraction, museum, 
monument, art, castle, fort, arts\_centre, 
library, community\_centre, ruins, 
pofw (place of worship), hairdresser & NA & 20 \\ \hline
Shop & OSM & convenience, pharmacy, clothes, florist, 
chemist, bookshop, butcher, shoe\_shop, 
beverages, jeweller, gift\_shop, 
sports\_shop, stationery, 
outdoor\_shop, mobile\_phone\_shop, 
toy\_shop, newsagent, bakery, 
kiosk, greengrocer, beauty\_shop, 
video\_shop, computer\_shop, bicycle\_shop, 
laundry, furniture\_shop, post\_office, 
bank, supermarket, mall, 
department\_store, garden\_centre, 
car\_dealership, doityourself & NA & 20 \\ \hline
Hospitality & OSM & cafe, food\_court, bar, restaurant, 
biergarten, pub, fast\_food & NA & 20 \\ \hline
Outdoor & FSO & NA & Land use classified as ‘Green spaces and recreation areas’, 
‘Alpine pastures’, ‘Lake and rivers’, and ‘Forests’ & 35 \\ \hline
Hotel & OSM & hotel, motel, hostel, 
camp\_site, caravan\_site, 
alpine\_hut, bed\_and\_breakfast,
guesthouse, chalet & NA & NA \\ \hline
Other Home & FSO & NA & Residential population & 20 \\ \hline
Healthcare & OSM & hospital, doctors, dentist, veterinary, optician, clinic & NA & \\ \hline
University & OSM & university, École Polytechnique Fédérale de Lausanne, 
Université de Lausanne, Haute école, 
École hôtelière, Université, Hotel Management, 
UNIL, EPFL, HESAV, HEG-GE, HES-SO, UNIGE (university names) & NA & NA \\ \hline
Primary School & Cantonal offices & NA & School locations & 35 \\ \hline
Secondary School & Cantonal offices & NA & School locations & 35 \\ \hline

\end{longtable}

FSO: Federal Statistical Office \cite{noauthor_population_2021, noauthor_swiss_2021, noauthor_swiss_2021-1}; OSM: OpenStreetMap \cite{noauthor_openstreetmap_2024}.

\setcounter{figure}{0}

\begin{figure}[!h]
\includegraphics[width=0.9\textwidth]{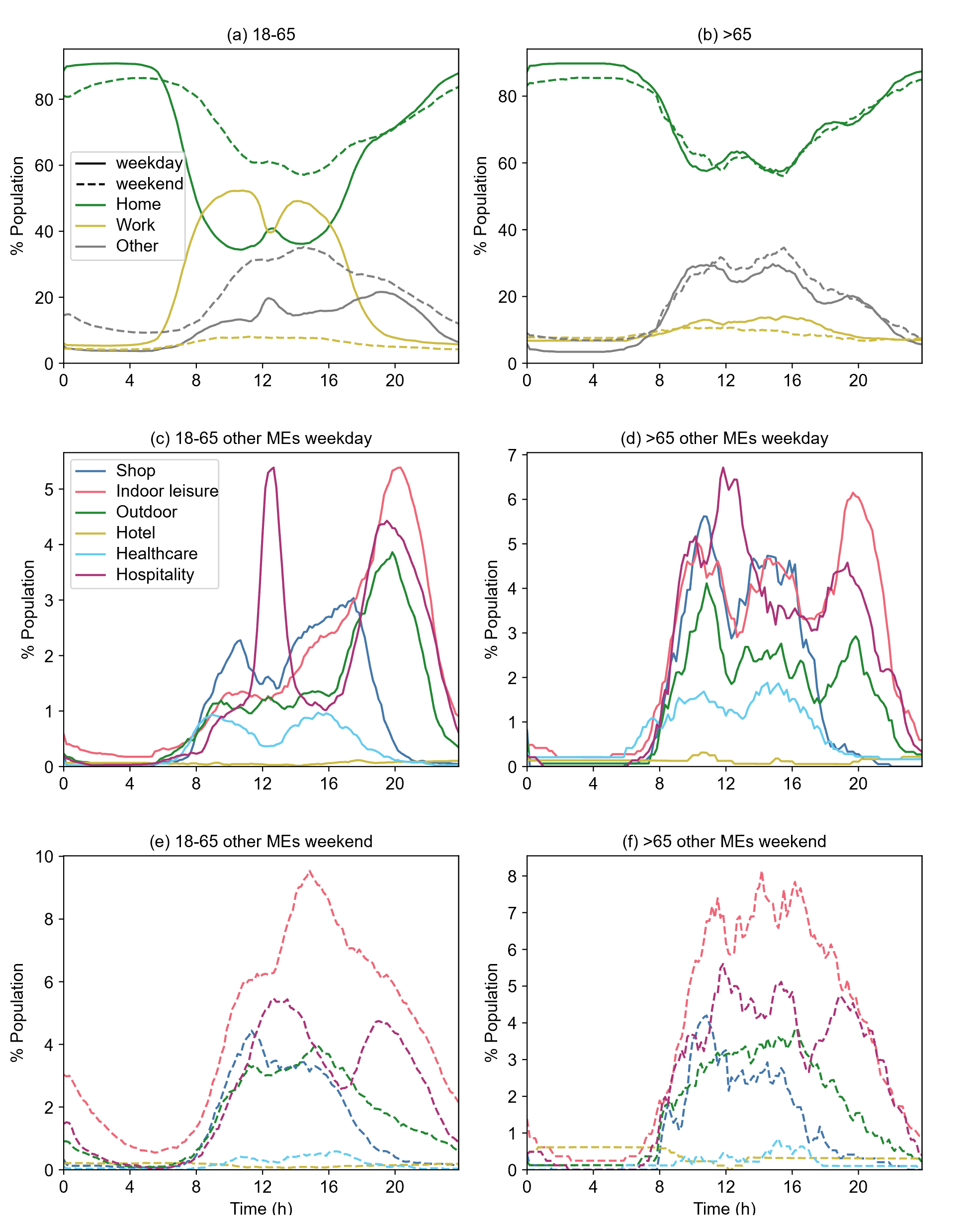} 
\centering
\caption{{\bf Diurnal ME occupancy profiles for different age group totals processed from TimeUse+.} Population of both Cantons in different MEs at 10-min resolution for (a,c,e) adults (age 18-65) and (b,d,f) seniors ($>$65). (a,b) profiles for `Home', `Work' and combined remaining MEs (`Other') for both day types; details of remaining MEs for (c,d) weekday and (e,f) weekend days.}
\label{sfig0}
\end{figure}

\begin{figure}[!h]
\includegraphics[width=0.9\textwidth]{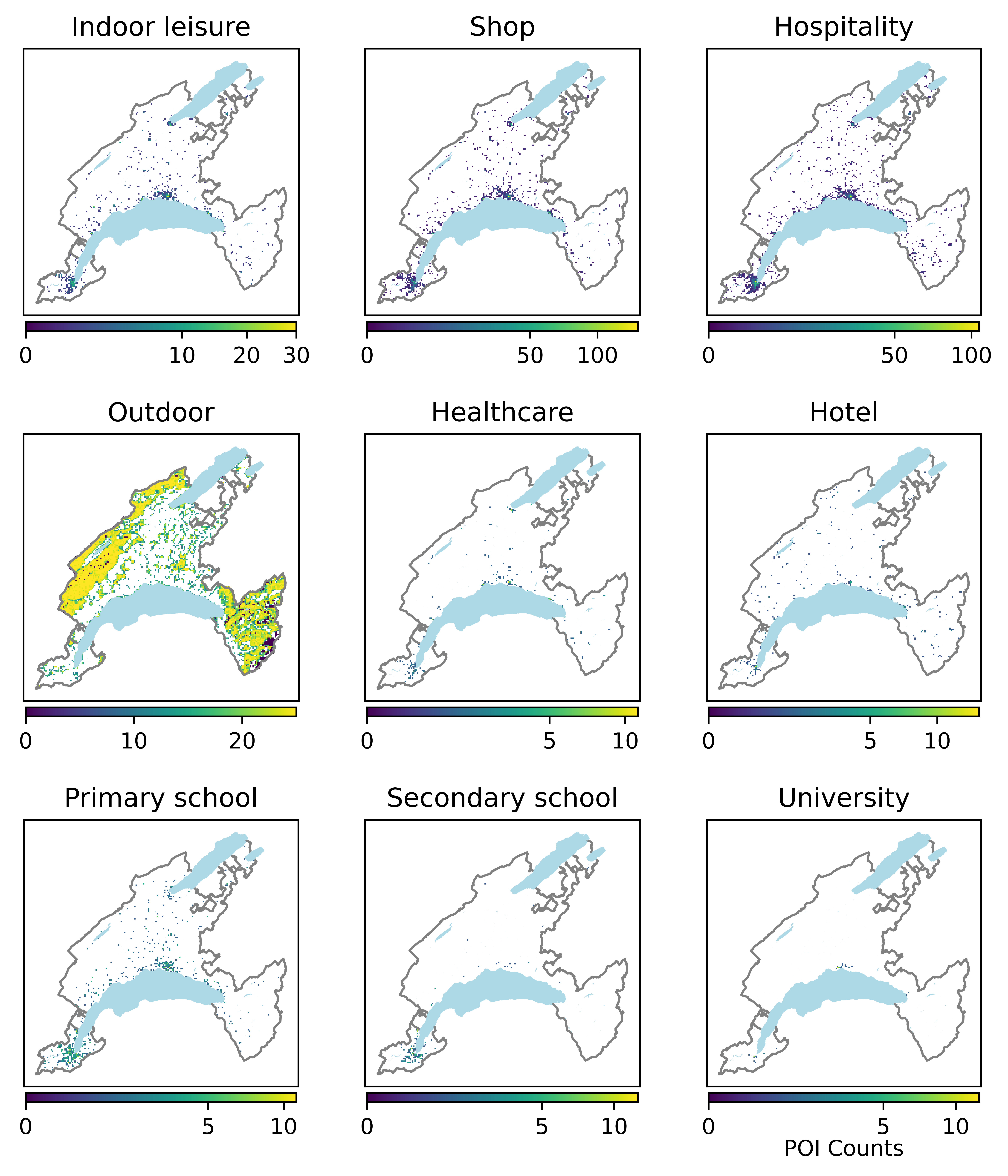} 
\centering
\caption{{\bf Point-Of-Interest counts for microenvironments in each 500~m grid neighbourhood.} The counts for Outdoor ME show the number of 100 $m^2$ green or water-side area in the 500 m grid neighbourhoods. The colour scale is normalized by a power law with exponent of 0.5 using the colors.PowerNorm function from the Python package Matplotlib except for Outdoor ME which uses a linear colour scale.}
\label{sfig1}
\end{figure}

\begin{figure}[!h]
\includegraphics[width=0.5\textwidth]{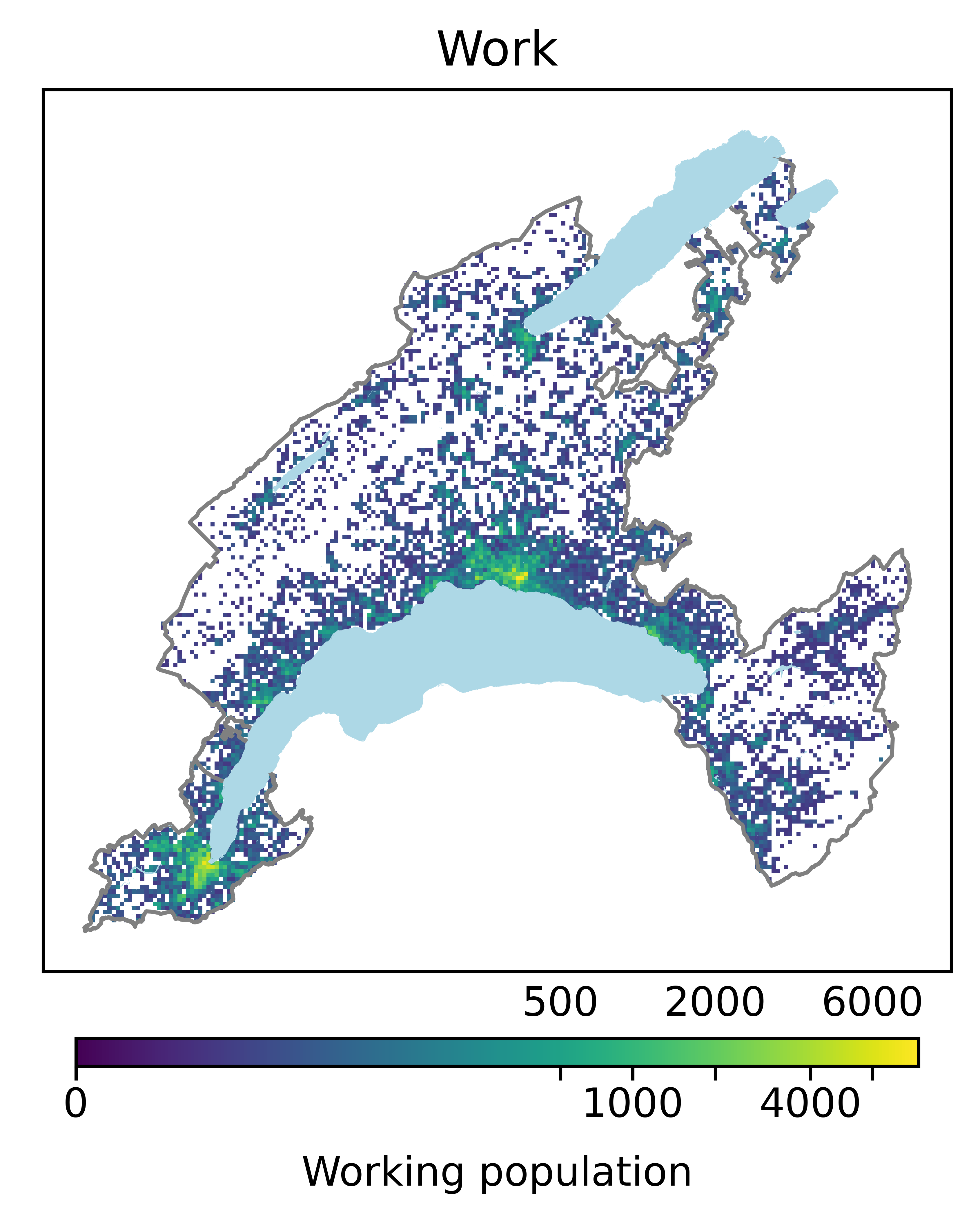} 
\centering
\caption{{\bf Work population in each 500~m grid cell neighbourhood.} Processed from census data \cite{noauthor_employed_2020}. The colour scale is normalized by a power law with exponent of 0.2 using the colors.PowerNorm function from the Python package Matplotlib.}
\label{sfig2}
\end{figure}

\begin{figure}[!h]
\includegraphics[width=0.5\textwidth]{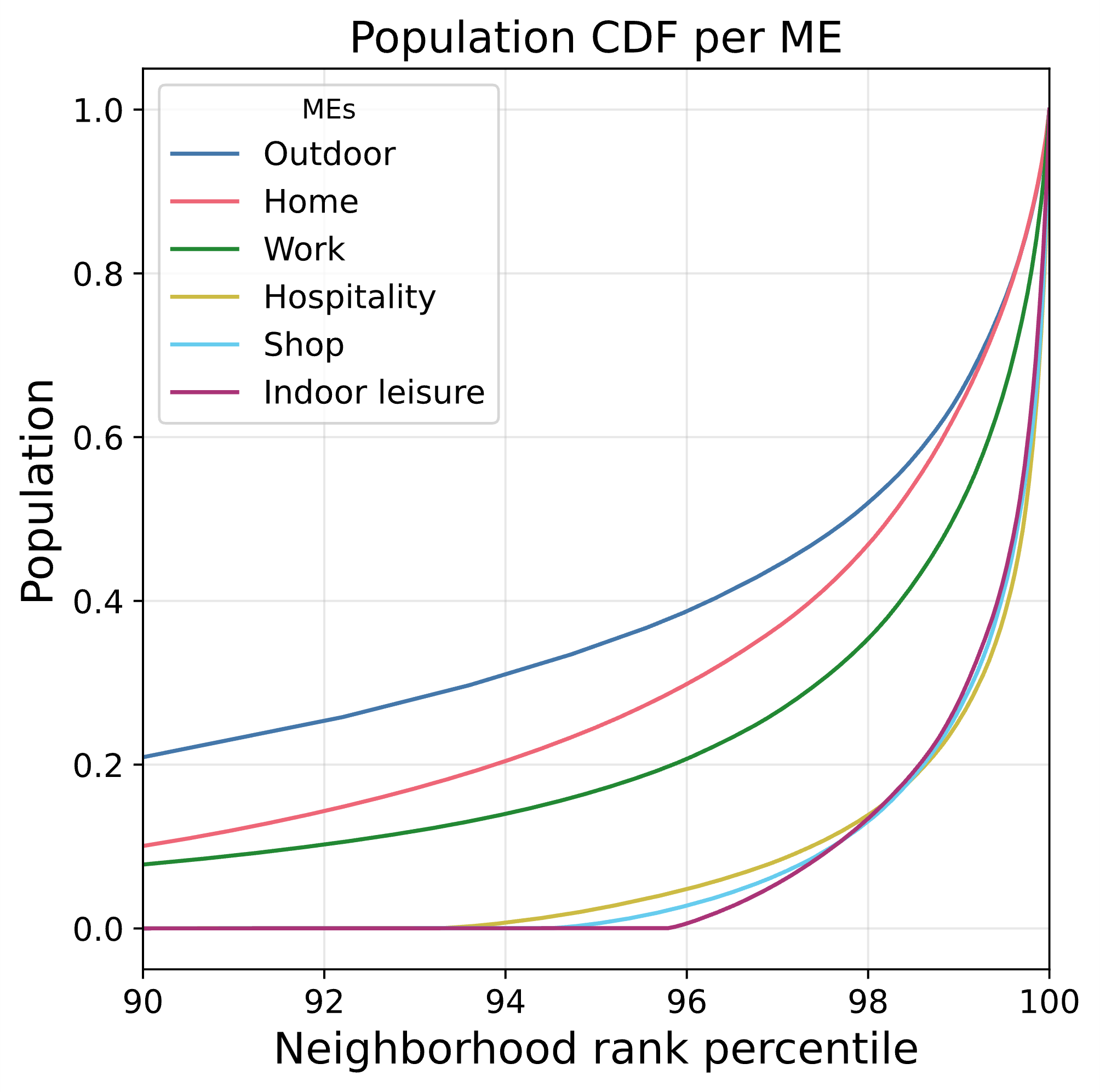} 
\centering
\caption{{\bf Cumulative probability distribution of population in selected microenvironments.}
Neighbourhoods (500~m grid cells) are ranked by the population in a specific ME from low to high over the whole simulation domain (x axis) with cumulative population to the total population ratio in that ME (y axis).}
\label{sfig3}
\end{figure}

\begin{figure}[!h]
\includegraphics[width=0.9\textwidth]{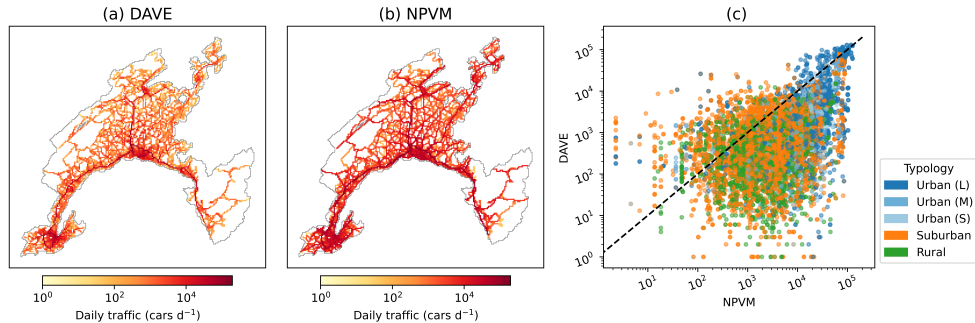} 
\centering
\caption{{\bf Comparison of DAVE and NPVM modelled traffic flows on a weekday.} NPVM road traffic flows are accessed from \cite{noauthor_load_2025} for the year 2017 and aggregated to 500~m grid cells used in DAVE modelling by taking the maximum road-based values in the grid cell and adjusting by $y=2.28x$, as derived in Fig.~\ref{sfig5}. As NPVM does not include every road, only grid cells where NPVM has values are shown. (c) Traffic flow comparison with the typology of each grid cell classified based on \cite{noauthor_spatial_2020}. Due to uncertainty in the aggregation process, this comparison is to evaluate full-domain spatial patterns (a, b) rather than to verify exact traffic values (c). To evaluate traffic values per grid cell, it is more accurate to directly compare with the SARTC measurements (Fig.~\ref{fig6})}
\label{sfig4}
\end{figure}

\begin{figure}[!h]
\includegraphics[width=0.9\textwidth]{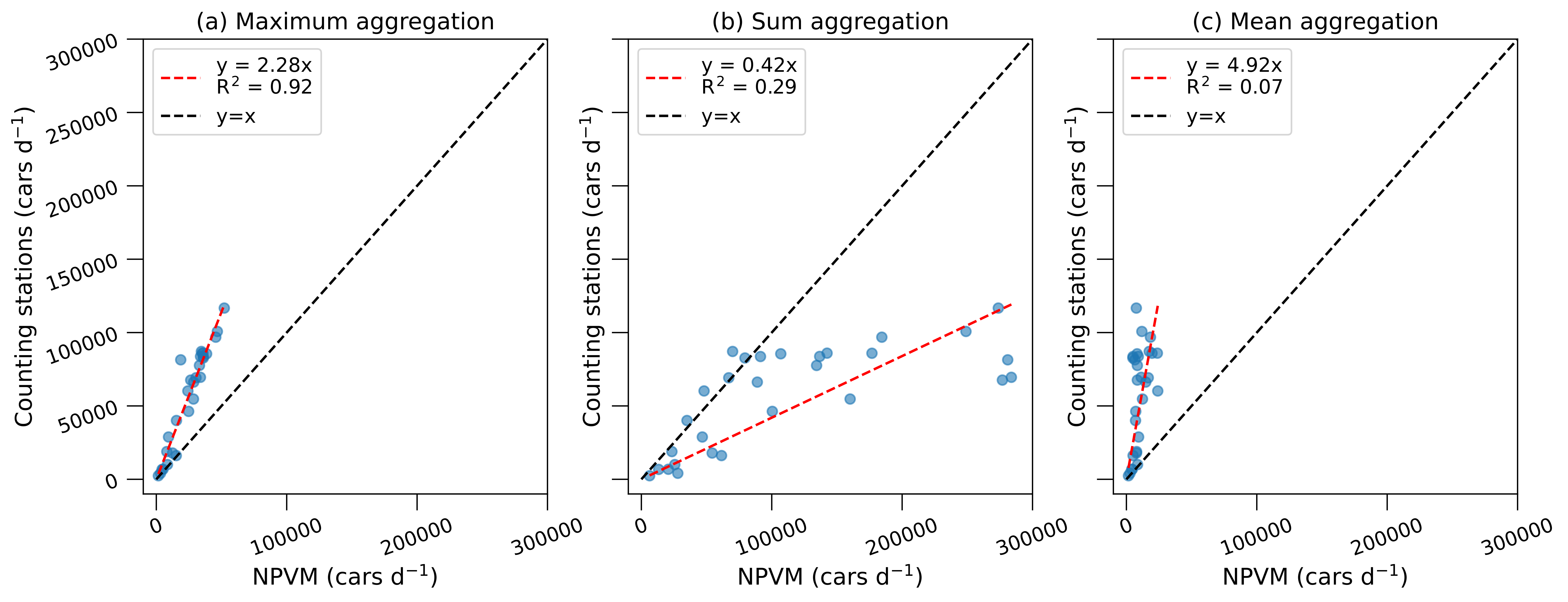} 
\centering
\caption{{\bf Comparison between NPVM traffic flows aggregated by different methods with the automatic counts from SARTC.} NPVM (2017) \cite{noauthor_load_2025} road traffic flows aggregated to 500~m grid cells using three methods: (a) maximum, (b) sum, or (c) mean of the road-based values in the grid cell. The aggregated values are compared with automatic counts from SARTC (2022) \cite{noauthor_swiss_2022}, which are considered a good approximation of grid-based travel flows here because they are placed at critical transport locations and include both directions. The linear fits between NPVM aggregated values and SARTC are shown for each method. Taking maximum and adjusting with the linear fit can provide reliable estimations of grid-based travel flows (a).}
\label{sfig5}
\end{figure}

\begin{figure}[!h]
\includegraphics[width=0.8\textwidth]{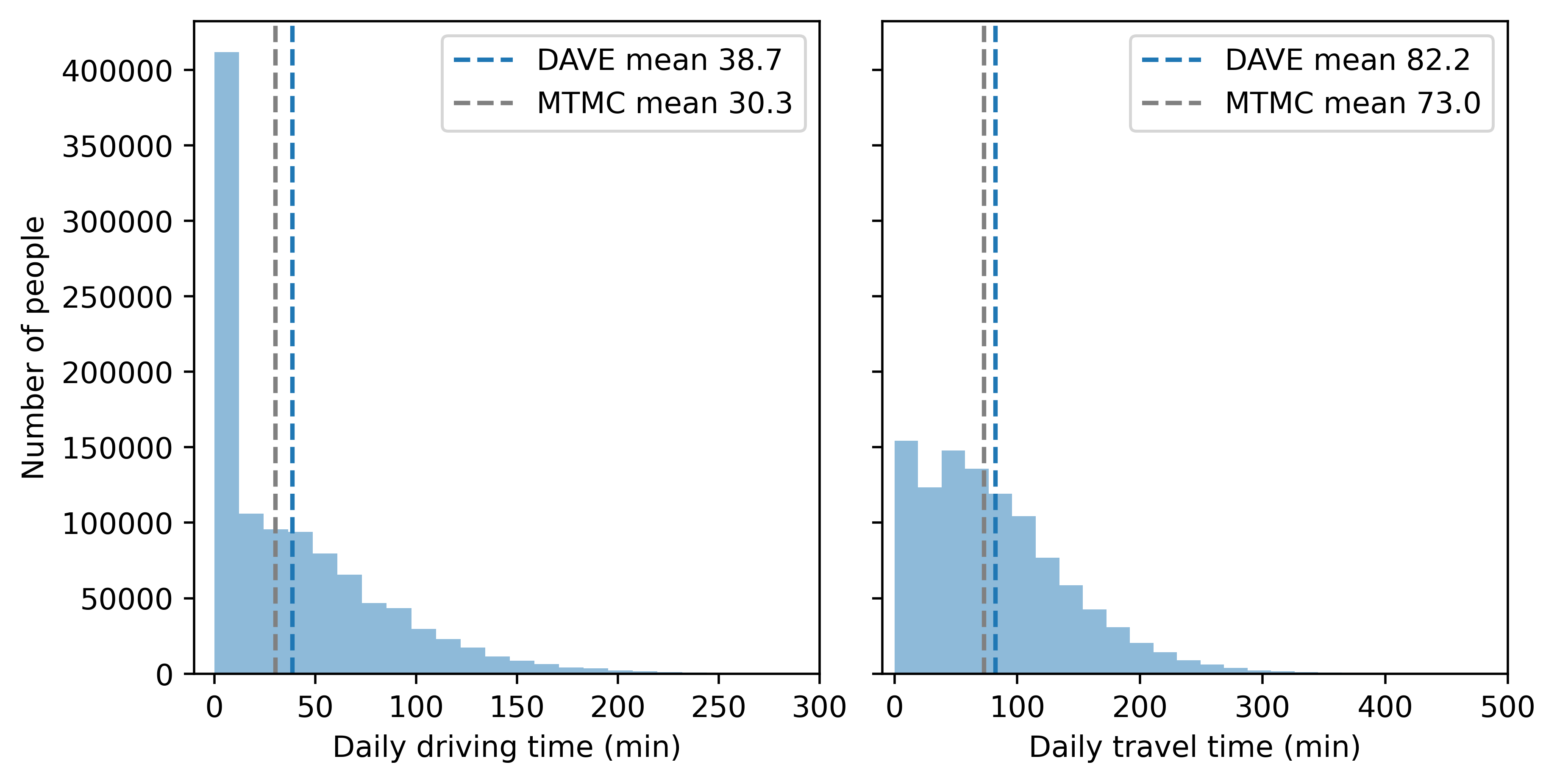} 
\centering
\caption{{\bf Probability distribution of daily driving and total travel time on a weekday.}}
\label{sfig6}
\end{figure}

% Include only the SI item label in the paragraph heading. Use the \nameref{label} command to cite SI items in the text.
% \paragraph*{S1 Fig. }
% \label{S1_Fig}
% {\bf Bold the title sentence.} Add descriptive text after the title of the item (optional).

\end{document}